\documentclass[aps,reprint,twocolumn,notitlepage]{revtex4-1}
\usepackage{graphicx}
\usepackage[T1]{fontenc}
\usepackage{amsmath}
\usepackage[utf8]{inputenc}
\usepackage[usenames,dvipsnames]{xcolor}

\newcommand{\ua}{\textbf{u}_{\text{A}}}
\newcommand{\da}{\boldsymbol\nabla\cdot\textbf{u}_{\text{A}}}
\newcommand{\dap}{\boldsymbol\nabla'\cdot\textbf{u}_{\text{A}}'}
\newcommand{\uh}{\textbf{u}}
\newcommand{\dv}{\boldsymbol\nabla\cdot\textbf{u}}
\newcommand{\dvp}{\boldsymbol\nabla'\cdot\textbf{u}'}

\begin{document}

\title{Energy cascade rate in isothermal compressible magnetohydrodynamic turbulence}

\author{N. Andr\'es$^{1}$} \author{F. Sahraoui$^{1}$} \author{S.
Galtier$^{1,2}$} \author{L. Z. Hadid$^{3}$} \author{P. Dmitruk$^{4,5}$}
\author{P. D. Mininni$^{4,5}$} \affiliation{$^1$ Laboratoire de Physique des Plasmas, \'Ecole Polytechnique, CNRS, Sorbonne University, Observatoire de Paris, Univ. Paris-Sud, F-91128 Palaiseau Cedex, France \\ $^2$ Univ. Paris-Sud, Universit\'e Paris-Saclay, France. \\ $^3$ Swedish Institute of Space Physics, Uppsala, Sweden.\\ $^4$ Departamento de F\'{\i}sica, Facultad de Ciencias Exactas y Naturales, Universidad de Buenos Aires, Ciudad Universitaria, 1428, Buenos Aires, Argentina.\\ $^5$ Instituto de F\'{\i}sica de Buenos Aires, CONICET-UBA, Ciudad Universitaria, 1428, Buenos Aires, Argentina.}

\begin{abstract}
Three-dimensional direct numerical simulations are used to study the energy cascade rate in isothermal compressible magnetohydrodynamic turbulence. Our analysis is guided by a two-point exact law derived recently for this problem in which flux, source, hybrid{,} and mixed terms are present. The relative importance of each term is studied for different initial subsonic Mach numbers $M_S$ and different magnetic guide fields ${\bf B}_0$. The dominant contribution to the energy cascade rate comes from the compressible flux, which depends weakly on the magnetic guide field ${\bf B}_0$, unlike the other terms whose modulus increase significantly with $M_S$ and {${\bf B}_0$}. In particular, for strong {${\bf B}_0$} the source and hybrid terms are dominant at small scales with almost the same amplitude but with a different sign. A statistical analysis made with {an isotropic} decomposition based on the SO(3) rotation group is shown to generate spurious results in presence of ${\bf B}_0${, when compared with an axisymmetric decomposition better suited to the geometry of the problem.} Our numerical results are compared with previous analyses made with in-situ measurements in the solar wind and the terrestrial magnetosheath.
\end{abstract}

\maketitle

%-------------------------------------------------------------
\section{Introduction}\label{intro}
%-------------------------------------------------------------

Exact results in fully developed turbulence represent strong boundary conditions that any model must satisfy \citep{F1995}, {even though} there are only a few of {such} predictions. {The so-called ``4/5 law" is an exact relation for incompressible hydrodynamic (HD) turbulence. In the infinite Reynolds number limit and assuming space homogeneity, isotropy{,} and time stationarity, this law expresses how the two-point third-order structure function for the velocity field is connected to the energy cascade rate $\varepsilon$. In particular, in Fourier space {and in the absence of intermittency,} this exact relation leads to the well-known Kolmogorov energy spectrum $E_k\sim \varepsilon^{2/3} k^{-5/3}$ \citep{K1941a,K1941b}.} For incompressible magnetohydrodynamic (IMHD) turbulence, {a first attempt at deriving such relations was done by} \citet{Ch1951}, under the assumptions of infinite kinetic and magnetic Reynolds numbers, time stationarity, space homogeneity, and full isotropy (i.e., rotation and mirror symmetries). Later, \citet{P1998a,P1998b} derived the so-called {\it 4/3 law} for IMHD turbulence, which gives a simple relation between two-point third-order structure functions, the distance between the two points{,} and the energy dissipation rate.

The validity of the exact law in IMHD turbulence has been the subject of several numerical tests \citep[see, e.g.][]{Mi2009,B2009,W2010,Y2012}. For example, \citet{Mi2009} reported high spatial resolution results for decaying IMHD turbulence in which the energy dissipation rate seem{s} to reach asymptotically a constant value at large Reynolds numbers. An extension of the exact IMHD law in presence of a constant velocity shear was proposed and tested numerically with direct numerical simulations (DNS{s}) of two-dimensional (2D) IMHD \citep{W2010}. Among several other uses, the exact laws for IMHD turbulence provide a precise identification of the inertial range \citep[see, {e.g.,}][and references therein]{A2016b}, {and} an estimate of the energy cascade rate and the Reynolds numbers in {experiments of turbulence}, in particular when dissipati{on} mechanisms are unknown {such as} in near-Earth space plasmas \citep{SV2007,WEY2007,M2008,Sa2008}.

Under the classical assumptions of homogeneity, stationarity{,} and infinite kinetic/magnetic Reynolds numbers, \citet{B2013} derived an exact law for isothermal compressible MHD (CMHD) turbulence. Their results revealed the presence of a new type of term that acts in the inertial range as a {\it source} (or a sink) for the energy cascade rate \citep[see also,][]{Ga2011}. It is worth noticing that in IMHD turbulence there is only one type of term, the flux, that transfers energy in the inertial range \citep{MY1975,F1995}. Because of its complexity, the expression of the exact law in CMHD is not unique \citep[e.g., see][]{BF2018}. For example, \citet{A2017b} have re-derived the law using the plasma velocity, the compressible Alfv\'en velocity{,} and the plasma density as primitive variables. The authors found four different categories of terms that are involved in the inertial range. Besides the flux and the sources previously reported, the authors {also} found two new types of terms to which they referred to as {\it hybrid} and $\beta$-{\it dependent} terms (with $\beta$ the ratio between the plasma and magnetic pressure). {One of the goals} of the present paper {is thus} to investigate numerically the relative importance {and the} contribution of {each of these terms to} the exact law in CMHD isothermal turbulence.

The role of density fluctuations in the solar wind energy cascade rate was investigated by \citet{C2009b}. Using Ulysses solar wind data the authors found a better scaling relation with a heuristic compressible model than with the IMHD exact relation, showing therefore the relevance of density fluctuations in the cascade process (see {a} discussion of this model in \citet{H2017a}). Following a more rigorous approach, \citet{B2016c} used the exact law for isothermal CMHD \citep{B2013} to analyze the fast solar wind data from the THEMIS mission. The authors performed a term-by-term analysis, showed the existence of an inertial range over more than two decades of scales, and found that the compressible fluctuations {increase} (from 2 to 4 times) the {estimation of the} turbulent cascade rate with respect to the {estimations stemming from the} incompressible model. \citet{H2017a} extended the previous analysis (still {using} THEMIS data) to the slow solar wind which is known to be more compressible. In {this case} they found that the compressible energy cascade rate is {increased even further (because of higher density fluctuations in the slow solar wind when compared to the fast wind)} and that it {obeys} a power-law scaling with the turbulent Mach number. {However, it is worth noticing} that in all these recent studies \citep{B2016c,H2017a,H2017b} several source terms of the exact CMHD law have been neglected. It is {another} goal of the present paper to check carefully if the assumptions made to neglect these terms are indeed satisfied in DNS close to the solar wind conditions.

Recently, several new results have been obtained in compressible turbulence that are worth mentioning here. For example, \citet{Z2017a} used the nearly incompressible MHD (NI MHD) equations \citep[e.g., see][]{Z1990} to describe solar wind homogeneous or inhomogeneous turbulence for plasma $\beta\lesssim 1$. The authors presented a NI MHD formulation describing the transport {throughout the solar wind of turbulence which was in its majority 2D, and with a small slab component}. Using Voyager 1 measurements, \citet{Z2017b} showed that inner heliosheath fast and slow MHD waves incident on the heliopause generate, in the very local interstellar medium (LISM), only fast MHD waves that propagate into this medium. The authors suggested that this may be the origin of compressible turbulence in the LISM.

On the other hand, \citet{Y2017} used DNS of mechanically forced CMHD turbulence to study the degree to which some turbulence theories proposed for incompressible flows remain applicable in the compressible case. In particular, intermittency, coherent structures{,} and energy cascade rates were studied with different forcing mechanisms. \citet{Grete2017} extended the classical shell-to-shell energy transfer analysis to the isothermal compressible regime. The authors derived four new transfer functions in order to measure{,} e.g.{,} the energy exchange via the magnetic pressure. \citet{A2017a} showed direct numerical evidence of the excitation of magnetosonic and Alfv\'en waves in three-dimensional (3D) CMHD turbulence at small sonic Mach numbers. Using spatio-temporal spectra, in the low $\beta$ regime, the authors found excitation of compressible and incompressible fluctuations, with a clear transfer of energy towards Alfv\'enic and 2D modes. However, in the high $\beta$ regime, fast and slow magnetosonic waves were present with no clear signature of Alfv\'en waves, a significant part of the energy being carried by 2D turbulent eddies. Finally, \citet{A2018a} derived an exact law for 3D homogeneous compressible isothermal Hall magnetohydrodynamic turbulence, without the assumption of isotropy. The authors showed that the Hall current introduces new flux and source terms that act at the small scales (comparable or smaller than the ion skin depth) to significantly impact the turbulence dynamics.

The main goal of the present paper is {thus} to investigate the energy cascade rate in isothermal CMHD turbulence using 3D DNS{s}. We present a comprehensive analysis of the exact law, with a particular emphasis on the nature of each term involved in the nonlinear cascade of energy{,} and on the role of the background magnetic field ${\bf B}_0$. Furthermore, we discuss our numerical results {in the context of} the original observational results from Refs. \citep{B2016c,H2017a}. We expect that our numerical findings will help to clarify some subtle issues regarding the use of the compressible exact law in DNS{s} and spacecraft data.

The paper is organized as follows: in Sec.~\ref{equations} we describe the CMHD equations; in Sec.~\ref{exact} we present the exact law for fully developed isothermal CMHD turbulence; in Sec.~\ref{numcode} and \ref{corrfun} we introduce the numerical code and techniques used to compute the different correlation functions; in Sec.~\ref{results} we expose our numerical results and, finally, in Sec.~\ref{discussion} we discuss the main findings and their implications for the observational studies in the near-Earth space.

%-------------------------------------------------------------
\section{Theory}
%-------------------------------------------------------------

%-------------------------------------------------------------
\subsection{Compressible MHD}\label{equations}
%-------------------------------------------------------------

The 3D CMHD equations correspond to the continuity equation for the mass density, the momentum equation for the velocity field in which the Lorentz force is included, the induction equation for the magnetic field, and the differential Gauss' law.  These equations can be written as \citep[see, e.g.,][]{F2014,A2017a},
\begin{align}\label{modeld:1} 
    & \frac{\partial \rho}{\partial t}=-\boldsymbol\nabla\cdot(\rho \textbf{u}) , \\ \label{modeld:2}
    & \frac{\partial
	\textbf{u}}{\partial t} = -\textbf{u}\cdot\boldsymbol\nabla\textbf{u}-\frac{\boldsymbol\nabla P}{\rho} + \frac{(\boldsymbol\nabla\times\textbf{B})\times\textbf{B}}{4\pi\rho} + \textbf{f}_k+\textbf{d}_k , \\ \label{modeld:3}
	& \frac{\partial \textbf{B}}{\partial t} = \boldsymbol\nabla\times\left(\textbf{u}\times\textbf{B}\right) + \textbf{f}_m + \textbf{d}_m , \\ \label{modeld:4}
	&\boldsymbol\nabla\cdot\textbf{B}=0 ,
\end{align}
where \textbf{u} is the velocity field fluctuation, $\textbf{B}=\textbf{B}_0+\textbf{b}$ is the total magnetic field, $\rho$ is the mass density{,} and $P$ is the scalar pressure. For the sake of simplicity we assume that the plasma follows an isothermal equation of state, $P=c_s^2\rho$, where $c_s$ is the constant sound speed, {which allows us to close the hierarchy of the fluid equations (no energy equation is further needed).} Finally, \textbf{f}$_{k,m}$ are {respectively} a mechanical and {the curl of the} electromotive large-scale forcing{s,} and $\textbf{d}_{k,m}$ are {respectively} the small-scale kinetic and magnetic dissipation terms.

Alternatively to the magnetic field \textbf{B}, the compressible Alfv\'en velocity $\ua\equiv\textbf{B}/\sqrt{4\pi\rho}$ can be used (where $\rho$ is time and space dependent). In this manner, both field variables, $\uh$ and $\ua${,} are expressed in speed unit{s}. Therefore, Eqs. \eqref{modeld:1}-\eqref{modeld:4} can be cast as \citep{M1987},
\begin{align}\label{1} 
    & \frac{\partial e}{\partial t}  = - \uh\cdot\boldsymbol\nabla e-c_s^2\boldsymbol\nabla\cdot\uh , \\ \nonumber
	&\frac{\partial \textbf{u}}{\partial t} = -\uh\cdot\boldsymbol\nabla\uh  + \ua\cdot\boldsymbol\nabla\ua - \frac{1}{\rho}\boldsymbol\nabla(P+P_M)  \\  \label{2} 
    & ~~~~~~~ - \ua(\da) + \textbf{f}_k  + \textbf{d}_k , \\ 	\label{3} 
    &\frac{\partial \ua}{\partial t} = - \uh\cdot\boldsymbol\nabla\ua + \ua\cdot\boldsymbol\nabla\uh -\frac{\ua}{2}(\dv) + \textbf{f}_m + \textbf{d}_m , \\  \label{4} 
    &\ua\cdot\boldsymbol\nabla\rho = -2\rho(\da),
\end{align}
where $P_M\equiv\rho u_\text{A}^2/2$ is the magnetic pressure. Note that we have written Eq. \eqref{modeld:3} as a function of the internal compressible energy for an isothermal plasma, i.e.{,} $e \equiv c_s^2\ln(\rho/\rho_0)$, where $\rho_0$ is a constant (of reference) mass density. {In the rest of the paper we shall assume that the fields considered are regular and therefore differentiable. Singular fields may exist in the inviscid case, leading to the appearance of anomalous dissipation \citep{DR2000,E2018,Ga2018}.}

%-------------------------------------------------------------
\subsection{Exact law for CMHD turbulence}\label{exact}
%-------------------------------------------------------------

Following the usual assumptions for fully developed homogeneous turbulence (i.e., infinite kinetic and magnetic Reynolds numbers and a steady state with a balance between forcing and dissipation \citep{Ga2011,B2013,A2016b,A2016c,B2018}), an exact law for CMHD turbulence can be obtained as \citep{A2017b,B2013}, 
\begin{widetext}
	\begin{align}\nonumber
		-2\varepsilon_C =& \frac{1}{2}\boldsymbol\nabla_{\boldsymbol\ell}\cdot\big\langle [(\delta(\rho\uh)\cdot\delta\uh+\delta(\rho\ua)\cdot\delta\ua + 2\delta e\delta\rho\big]\delta\uh - [\delta(\rho\uh)\cdot\delta\ua+\delta\uh\cdot\delta(\rho\ua)]\delta\ua \big\rangle \\ \nonumber
	&+\langle[R_E'-\frac{1}{2}(R_B'+R_B)-E'+\frac{P_M'-P'}{2}](\dv)+[R_E-\frac{1}{2}(R_B+R_B')-E+\frac{P_M-P}{2}](\dvp)\rangle \\ \nonumber
	&+\langle[(R_H-R_H')-\bar{\rho}(\uh'\cdot\ua)+H'](\da)+[(R_H'-R_H)-\bar{\rho}(\uh\cdot\ua')+H](\dap)\rangle \\\nonumber
	& +\frac{1}{2}\langle\big(e'+\frac{u_\text{A}}{2}^{'2}\big)\big[\boldsymbol\nabla\cdot(\rho\uh)\big]+\big(e+\frac{u_\text{A}}{2}^2\big)\big[\boldsymbol\nabla'\cdot(\rho'\uh')\big]\rangle \\ \label{exactlaw}
	&-\frac{1}{2}\langle\beta^{-1'}\boldsymbol\nabla'\cdot(e'\rho\uh) + \beta^{-1}\boldsymbol\nabla\cdot(e\rho'\uh') \rangle ,
	\end{align}
\end{widetext}
where $\varepsilon_C$ is the total compressible energy cascade rate. We have defined the total energy and the density-weighted cross-helicity {per unit volume} respectively as
\begin{align}\label{energy}
	E(\textbf{x}) \equiv &~\frac{\rho}{2}(\uh\cdot\uh+\ua\cdot\ua) + \rho e , \\
	H(\textbf{x}) \equiv &~\rho(\uh\cdot\ua),
\end{align}
and their associated two-point correlation functions as,
\begin{align}
	R_E(\textbf{x},\textbf{x}') \equiv&~ \frac{\rho}{2}(\uh\cdot\uh'+\ua\cdot\ua') + \rho e'  ,\\
	R_H(\textbf{x},\textbf{x}') \equiv&~ \frac{\rho}{2}(\uh\cdot\ua'+\ua\cdot\uh') .
\end{align}
In addition, we have defined the magnetic energy density {as} $R_B(\textbf{x},\textbf{x}') \equiv \rho(\ua\cdot\ua')/2${. In all cases} the prime denotes field evaluation at $\textbf{x}'=\textbf{x}+\boldsymbol\ell$ ($\boldsymbol\ell$ being the displacement vector) and the angular bracket $\langle\cdot\rangle$ denotes an ensemble average. It is worth mentioning that the properties of spatial homogeneity implies (assuming ergodicity) that the results of averaging over a large number of realizations can be obtained equally well by averaging over a large region of space for one realization \citep{Ba1953}. Finally, we have introduced the usual increments and local mean definitions, i.e.{,} $\delta\alpha\equiv\alpha'-\alpha$ and $\bar{\alpha}\equiv(\alpha'+\alpha)/2$ (with $\alpha$ any scalar {or vector} function), respectively.

We recall that the derivation of the exact law \eqref{exactlaw} does not require the assumption of isotropy and that it is independent of the dissipation mechanisms {acting} in the plasma (assuming that the dissipation acts only at the smallest scales in the system) \citep[see also,][]{Ga2011,A2016b,A2016c}. {In a compact form, the exact law for CMHD turbulence (i.e., Eq. \ref{exactlaw}) can be schematically written as,
\begin{equation}\label{summarylaw}
	-2\varepsilon_C=\frac{1}{2}\boldsymbol\nabla_\ell\cdot\textbf{F}_\text{C}+\text{S}_\text{C}+\text{S}_\text{H}+\text{M}_\beta,
\end{equation}
where $\textbf{F}_\text{C}$, $\text{S}_\text{C}$, $\text{S}_\text{H}$ and M$_\beta$ represent the total compressible flux, source, hybrid and $\beta$-dependent terms, respectively, {which are defined as}
\begin{widetext}
	\begin{align}\label{term_flux}
		\textbf{F}_\text{C} \equiv&~ \langle [(\delta(\rho\uh)\cdot\delta\uh+\delta(\rho\ua)\cdot\delta\ua + 2\delta e\delta\rho\big]\delta\uh - [\delta(\rho\uh)\cdot\delta\ua+\delta\uh\cdot\delta(\rho\ua)]\delta\ua\rangle, \\ \nonumber
	  \text{S}_\text{C} \equiv&~ \langle[R_E'-\frac{1}{2}(R_B'+R_B)](\dv)+[R_E-\frac{1}{2}(R_B+R_B')](\dvp)\rangle \\ \label{term_source}
		&+\langle[(R_H-R_H')-\bar{\rho}(\uh'\cdot\ua)](\da)+[(R_H'-R_H)-\bar{\rho}(\uh\cdot\ua')](\dap)\rangle, \\ \nonumber
	  \text{S}_\text{H} \equiv&~ \langle\big(\frac{P_M'-P'}{2}-E'\big)(\dv)+\big(\frac{P_M-P}{2}-E\big)(\dvp)\rangle  + \langle H'(\da)+H(\dap)\rangle \\ \label{term_hybrid} &+\frac{1}{2}\langle\big(e'+\frac{u_\text{A}}{2}^{'2}\big)\big[\boldsymbol\nabla\cdot(\rho\uh)\big]+\big(e+\frac{u_\text{A}}{2}^2\big)\big[\boldsymbol\nabla'\cdot(\rho'\uh')\big]\rangle, \\ \label{term_beta}
		\text{M}_\beta \equiv&~ -\frac{1}{2}\langle\beta^{-1'}\boldsymbol\nabla'\cdot(e'\rho\uh) + \beta^{-1}\boldsymbol\nabla\cdot(e\rho'\uh') \rangle.
	\end{align}
\end{widetext}
}

{The quantity in eq.~\eqref{term_flux} is associated with the energy flux, and} {is the usual term present in the exact law of incompressible turbulence \citep{A2017b}. This term is written as a global divergence of products of increments of different variables. It is worth mentioning that the total compressible flux \eqref{term_flux} is a combination of fourth- and third-order terms, which makes a major difference with the incompressible case where the flux terms are usually third-order correlation functions. The {occurrence} of a fourth-order correlation function is a direct consequence of the total energy definition in the CMHD model (see Eq. \ref{energy}){, which is cubic in the fields}. The purely compressible source terms in Eq. \eqref{term_source}  may act as a source (or a sink) for the mean energy cascade rate in the inertial range. These terms involve two-point correlation functions (namely $R_E$, $R_B$ and $R_H$) and are proportional to the divergence of the Alfv\'en and kinetic velocity fields.}

{The hybrid term offers the freedom to be written either as a flux- or as a source-like term. However, when written as a flux-like term it cannot be expressed as the product of increments, as the usual flux in incompressible HD and MHD turbulence \citep{vkh1938,K1941a,K1941b,Ch1951,P1998a,P1998b} or the flux term in Eq.~{(\ref{term_flux})}. On the other hand, the mixed $\beta$-dependent term (already reported as a flux-like term in \citet{B2013} under {certain} condition{s}) has no counterpart in compressible HD turbulence \citep{Ga2011,B2014} and cannot, in general, be expressed either as purely flux or source. {Note also that} the mixed $\beta$-dependent term stems from the magnetic pressure gradient term in the momentum {Eq.~}\eqref{modeld:2}.}

{The schematic representation {in Eq.~}\eqref{summarylaw} {thus} reflects the nature of each term in the exact law for CMHD turbulence \citep{A2017b}{, and helps us quantify} the impact of each contribution {to} the nonlinear energy cascade rate. It is worth mentioning that in the observational works {in} Refs. \citep{B2016c,H2017a}, $\text{F}_\text{C}$, part of $\text{M}_\beta$ (under the assumption of statistical stationarity of the $\beta$ parameter), and part of $\text{S}_\text{H}$ were considered in the evaluation of the solar wind energy cascade rate. The remaining terms were considered as sources and assumed to be sub-dominant in the inertial range \citep[see,][]{K2013}. We will return to this issue in Sec. \ref{discussion}.}

{Under a sufficiently strong guide field $\boldsymbol{B}_0$, and assuming that the flow is statistically axisymmetric and has a weak dependence along the direction of the guide field, we can integrate Eq.~}{\eqref{summarylaw} over a cylinder of radius $\ell_\perp$ {and} obtain an approximate scalar relation for anisotropic turbulence in symbolic form (see also, Sec. \ref{corrfun}),
\begin{equation}\label{numlaw}
	-4\varepsilon_C\ell_\perp=\text{F}_\text{C}+\text{Q}_{\text{S}_\text{C}}+\text{Q}_{\text{S}_\text{H}}+\text{Q}_{\text{M}_\beta},
\end{equation}
where F$_\text{C}\equiv \textbf{F}_\text{C}\cdot\boldsymbol\ell_\perp/\ell_\perp$ and the integral functions correspond to
\begin{equation}
	\text{Q}_\text{T}\equiv\frac{2}{\ell_\perp}\int_0^{\ell_\perp}\text{T}(\ell_\perp^{*})\ell_\perp^{*} d\ell_\perp^{*} , 
\end{equation}
with $\text{T}(\ell_\perp)={\text{S}_\text{C}}(\ell_\perp)$, ${\text{S}_\text{H}}(\ell_\perp)$ and ${\text{M}_\beta}(\ell_\perp)$, respectively.}

%------------------------------------------------------------
\subsection{Numerical code}\label{numcode}
%-------------------------------------------------------------

\begin{table}{}\centering
	\begin{tabular}{p{.7cm} p{.7cm} p{.8cm} p{.8cm} p{.8cm}}
	\hline\hline
	Run & $B_0$ & $M_S$ & $\langle \delta E_u\rangle $ & $\langle \delta E_b\rangle$ \\
	\hline
		I  & 0 & $1/4$ & 0.13 & 0.14 \\
		II  & 2 & $1/4$ & 0.15 & 0.05 \\
	  III  & 8 & $1/4$ & 0.16 & 0.06 \\
		IV  & 0 & $1/2$ & 0.13 & 0.14 \\
	\hline\hline
	\end{tabular}
	\caption{Parameters used in Runs I to IV: $B_0$ is the magnetic guide field, $M_S$ is the sonic Mach number, $\langle \delta E_u\rangle$ and $\langle \delta E_B\rangle$ are the average fluctuating kinetic and magnetic energies reached in the stationary state.\label{table}}
\end{table}
The 3D CMHD Eqs. \eqref{modeld:1}-\eqref{modeld:4} are numerically solved using the Fourier pseudo-spectral code GHOST \citep{Go2005,Mi2011} with a new module for compressible flows based on previous developments \citep{Gh1993,D2005}. The numerical scheme used ensures the exact energy conservation for the continuous time spatially discrete equation{s} \citep{Mi2011} (as well as conservation of all other quadratic invariants in the system). We used {a linear} spatial resolution {of} $N=512$ {grid points in each direction} in a cubic periodic box. For simplicity, we used identical dimensionless viscosity and magnetic diffusivity, $\nu=\eta=1.25\times10^{-3}$ (i.e., the magnetic Prandtl number is {$P_m=\nu/\eta=1$}). {In all our runs, {the minimum wave number is} $k_{min} = 1$ for a box of length $L_0=2\pi${,} and $N=512$ lead{ing} to a maximum wavenumber $k_{max}=N/3~\approx~170$ ({resulting from the $2/3$} de-aliasing rule). At all times, we checked that $k_D/k_{max} < 1$,  $k_D$ being the dissipation wave number{, or in other words, that the simulations were well resolved.}}

The initial state of our simulations corresponds to density, velocity and magnetic fields amplitude fluctuations equal to zero. For all times $t>0$, {the velocity field and the magnetic vector potential are forced by a solenoidal} mechanical and an electromotive forcing, respectively, at the largest scales of the numerical box (i.e., {in the shell of modes in Fourier space with} $1\leq k_f \leq 3${, where $k_f$ are the forced wave numbers}). The mechanical and electromotive forcings are {random and} uncorrelated, and they inject neither kinetic nor magnetic helicity. {Furthermore, the set of random phases of the two forces are independent. {These r}andom phases {are} slowly evolv{ed} in time, to avoid introducing long-{time} correlations, but also to prevent introducing very fast {spurious} time scales. To this end, a new set of random phases is generated for each forcing function every 1/2 turnover time. Finally, the forcing{s} are linearly interpolated from their previous states to the new random states on 1/2 turnover time, and the process is then repeated (for {more} details about the random forcing {scheme} used here, see \citep{A2017a})}. We performed four numerical simulations with initial subsonic Mach numbers $M_S=u_0/c_s$ (tipically, $u_0{\approx}1$) and with different background magnetic field $B_0$ (see Table \ref{table}). This allows us to investigate different regimes of CMHD turbulence, with a special emphasis on the magnetic guide field and the level of compressibility of the plasma. In all cases studied here {\bf B}$_0$ is along the $\bf{\hat{z}}$ axis.

%------------------------------------------------------------
\subsection{Correlation functions}\label{corrfun}
%-------------------------------------------------------------

For the computation of correlation functions {in multiple directions (and thus to increase statistical convergence by averaging over all these directions)}, we {use} the angle-averaged technique presented in \citet{Ta2003}. This technique avoids the need to use 3D interpolations to compute the correlation functions in directions for which the evaluation points do not lie on grid points. This  significantly reduces the computational cost of {any} geometrical decomposition {of the flow} \citep{Ma2010}. In particular, and considering that we have simulations without and with a magnetic guide field, for which we can expect the fields to be respectively statistically isotropic or axisymmetric, we have used two decompositions: the one based on the SO(3) rotation group for isotropic turbulence, and another one based on the SO(2)$\times$R symmetry group (i.e., rotations in the $\hat{\textbf{x}}-\hat{\textbf{y}}$ plane plus translation{s} in the $\hat{{\bf z}}$ direction) for anisotropic (axisymmetric) turbulence. 

{The procedure used to average each term in Eq.~\eqref{exactlaw} over several directions can be summarized as follows: in the isotropic SO(3) decomposition, the correlation functions are computed along different directions generated by the vectors (all in units of grid points in the simulation box) (1,0,0), (1,1,0), (1,1,1), (2,1,0), (2,1,1), (2,2,1), (3,1,0), (3,1,1) and {those generated by} taking all the index and sign permutations of the three {spatial} coordinates (and removing any vector that is a positive or negative multiple of any other vector in the set) \citep{A1999,Ta2003}. This procedure generates 73 unique directions. {Field increments are then computed for all multiples of the 73 vectors.} In this manner, the SO(3) decomposition gives the correlation functions as a function of $73$ radial directions covering the sphere {in an approximately homogeneous way} \citep{Ta2003}, {and}  whose averaging results in the isotropic correlation functions that depend {solely} on $\ell$.}

{In the SO(2){$\times$R} case, the correlation functions are computed using 12 different directions generated by integer multiples of the vectors (1,0,0), (1,1,0), (2,1,0), (3,1,0), (0,1,0), (-1,1,0), (-1,2,0), (-2,1,0), (-1,2,0), (-1,3,0), (-3,1,0), (-1,3,0) ({as before}, all vectors are in units of grid points in the simulation box), and the vector (0,0,1) for the translations in {the} $z$ direction. Once all structure functions were calculated, the correlation functions are obtained by averaging over the 12 directions in the $\hat{\textbf{x}}-\hat{\textbf{y}}$ plane, and the parallel structure functions can be computed directly using the generator in the $\hat{{\bf z}}$ direction. In other words, the SO(2) decomposition gives the correlation functions along $12$ polar directions in the ${\bf \hat{x}}-{\bf \hat{y}}$ plane and after averaging, one obtains a final correlation function as a function of the perpendicular polar direction (i.e., $\ell_\perp$){,} while R {(the group of translations along ${\bf\hat{z}}$) is used to compute} the correlation function {in} the ${\bf\hat{z}}$ direction (i.e., with spatial increments $\ell_\parallel$) \citep{Im2017}.} {It is important to note that these two directions are not independent, and that if Eq.~\eqref{exactlaw} is used in this way to estimate parallel and perpendicular fluxes $\varepsilon_C^{(\parallel)}$ and $\varepsilon_C^{(\perp)}$, they won't be independent either. In practice, the SO(2) decomposition amounts to integrating Eq.~\eqref{exactlaw} over the surface of an infinite (or $2\pi$-periodic in our case) cylinder of radius $\ell_\perp$, under the assumption that for strong enough $B_0$ the fields are statistically axisymmetric and have weak dependence on the vertical (${\bf\hat{z}}$) coordinate, and thus $\ell_\perp$ increments dominate the structure functions with $\varepsilon_C^{(\perp)} \approx \varepsilon_C$. We can thus expect on one hand that this decomposition will give better estimations of the flux when $B_0$ is strong. On the other hand, as the directions used for the SO(2) decomposition are a subset of the 73 directions in the SO(3) decomposition, for zero or small $B_0$ the two decompositions can be expected to show similar scaling, albeit with poorer statistics in the former case. As will be shown in the next section, this is indeed the case.}

It is {also} worth mentioning that Eq.~\eqref{exactlaw} is {expected to hold} for the mean values and not for each particular direction. In each of these decompositions we thus average the $73$ (or $12$) correlation functions of each term in Eq. \eqref{exactlaw} to investigate their relative importance in the compressible energy cascade rate. Although the SO(3) decomposition is better suited for isotropic turbulence, it has been used before to investigate anisotropic turbulence for the analysis of experimental results \citep{K2000a,K2000b} and numerical simulations \citep{A1999,B2001,Mi2010,Im2011}. The SO(2)$\times$R decomposition, designed specifically from the symmetry group of axisymmetric turbulence, has been developed and used to investigate anisotropic turbulence using numerical simulations in \citep{Im2017}. In all cases, an improvement in the statistical convergence of correlation functions was observed when compared with correlation functions computed in only a few directions.

\begin{figure}\centering
	\includegraphics[width=.35\textwidth]{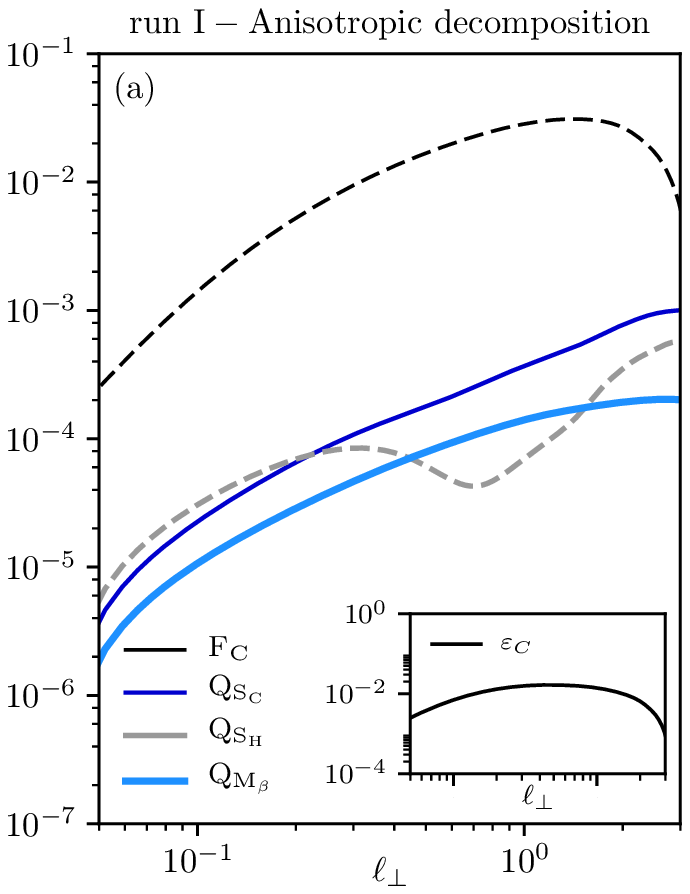} \\ \includegraphics[width=.35\textwidth]{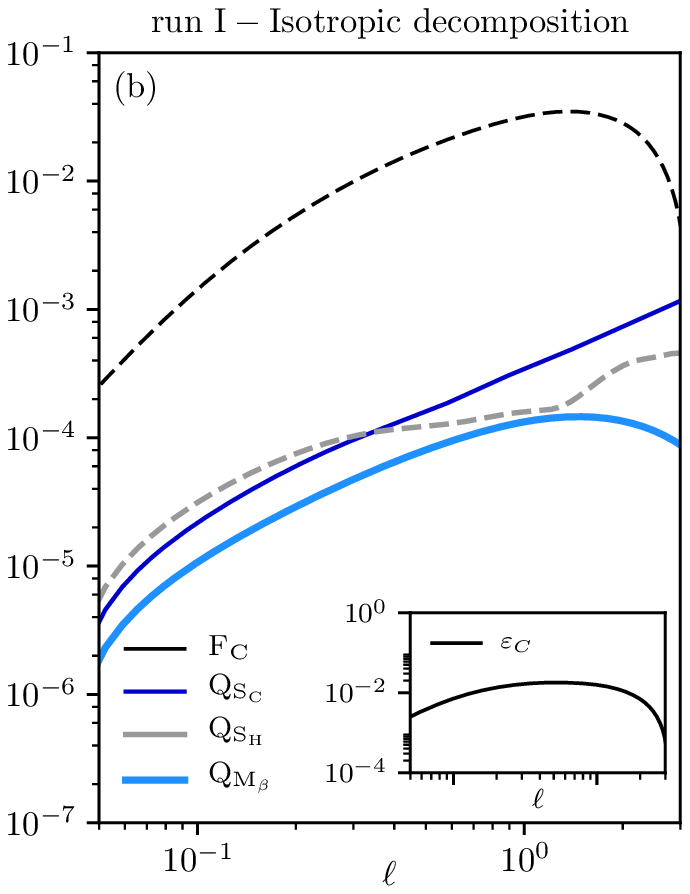}
	\caption{Run I: $B_0=0$ and $M_S=0.25$. Mean value of the compressible flux F$_\text{C}$ (black), source S$_\text{C}$ (dark blue), hybrid S$_\text{H}$ (gray) and $\beta$-dependent M$_\beta$ (light blue) terms of the exact law \eqref{numlaw} computed using the anisotropic (a) and isotropic (b) decompositions. Solid lines correspond to positive values while dashed lines correspond to negative values. Inset: total energy cascade rate computed using Eq. \eqref{numlaw}.}
	\label{fig1}
\end{figure}

\begin{figure}\centering
	\includegraphics[width=.35\textwidth]{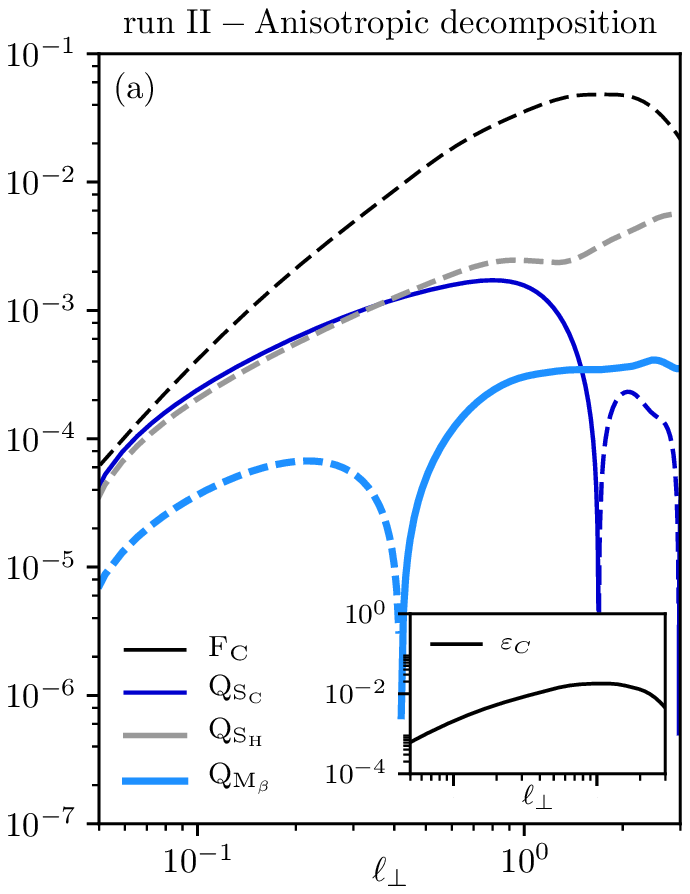} \\ \includegraphics[width=.35\textwidth]{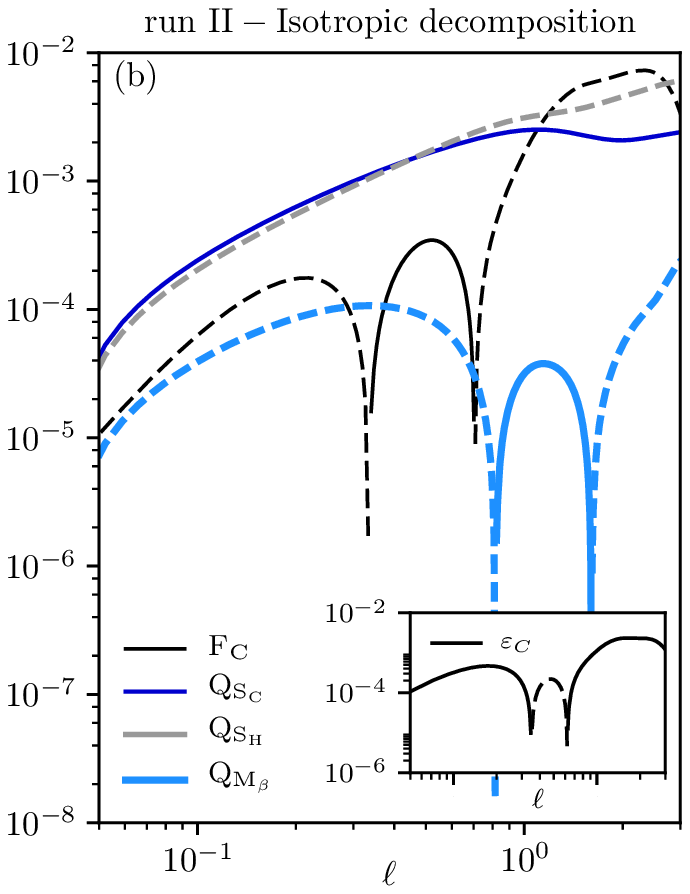}
	\caption{Run II: $B_0=2$ and $M_S=0.25$. Same description as in Fig.\,\ref{fig1} applies.}
	\label{fig2}
\end{figure}

\begin{figure}\centering
	\includegraphics[width=.35\textwidth]{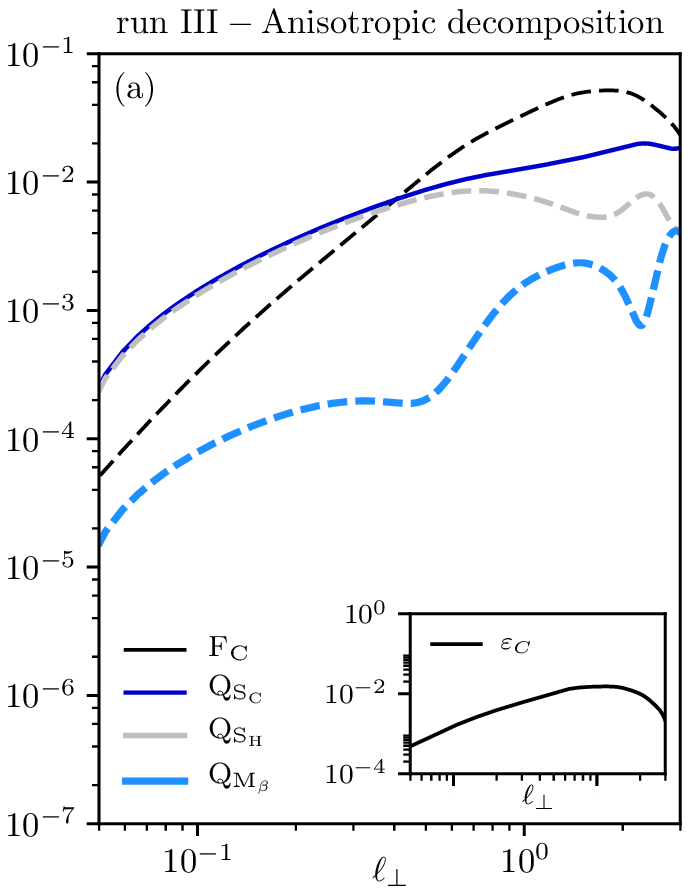} \\ \includegraphics[width=.35\textwidth]{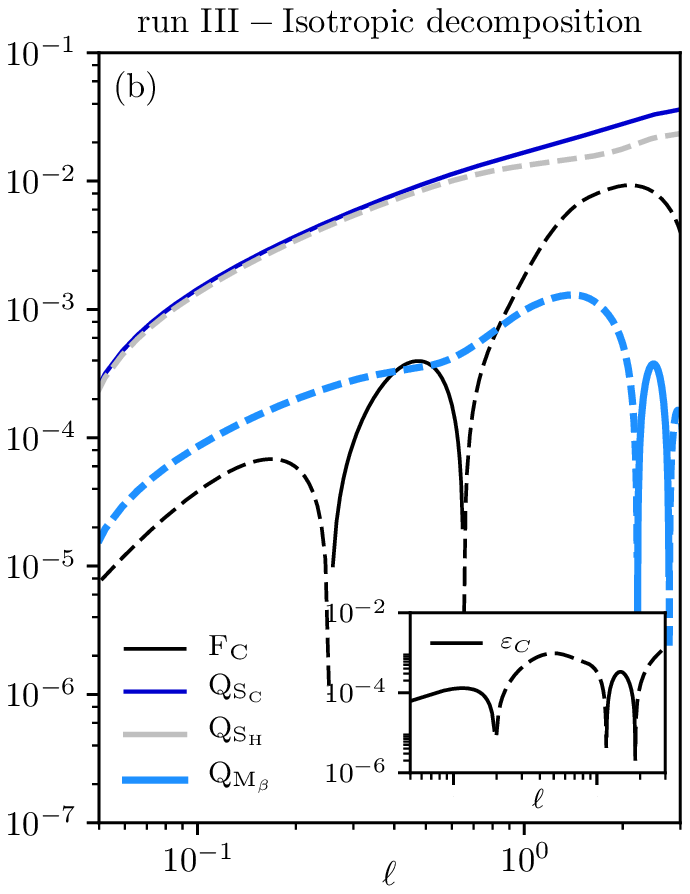}
	\caption{Run III: $B_0=8$ and $M_S=0.25$. Same description as in Fig.\,\ref{fig1} applies.}
	\label{fig3}
\end{figure}

%-------------------------------------------------------------
\section{Numerical results}\label{results}
%-------------------------------------------------------------

For all runs in Table \ref{table}, we computed the terms in the RHS of the exact law \eqref{exactlaw} using both the anisotropic and the isotropic decomposition techniques presented in Sec. \ref{numcode}. We investigate the different components and the energy cascade rate as we vary the sonic Mach number and the magnetic guide field in our simulations.

{Figures} \ref{fig1}(a) and \ref{fig1}(b) show for Run I the terms in the RHS of Eq. \eqref{numlaw} as a function of the perpendicular ($\ell_\perp$) and the isotropic ($\ell$) scale obtained using the anisotropic and isotropic decomposition, respectively. Since there is no privileged direction in Run I ({$B_0=|{\bf B}_0|=0$}), we find approximately the same variation and amplitude for the different terms as well as for the total energy cascade rate, independently of the decomposition used.

There are indications of a fully developed turbulence regime that is compatible with a Kolmogorov-like scaling \citep{K1941a,M1982,Mi2009,A2016b,Z2017a} and with a constant energy cascade rate {(see inset in Fig.~\ref{fig1})}. Note that at this moderate spatial resolution we cannot expect a wide inertial range. Nevertheless, the one evidenced here is sufficient for a first quantitative study of the different contributions to the exact law.

In the same format as Fig.~\ref{fig1}, Fig{s}.~\ref{fig2} and \ref{fig3} display the results for Runs II and III respectively. As expected, the presence of a magnetic guide field $B_0$ strongly affects the statistical results. First, the compressible flux decreases  slightly when $B_0$ is applied. We also see the appearance of a negative contribution (for Runs II and III) when the isotropic decomposition is used; this {disrupts} the scaling law that emerges. A comparison with the anisotropic decomposition reveals that {the disruptions are} a spurious effect due to the {assumption of isotropy}, which is not fulfilled in the runs with moderate to strong magnetic guide field \citep[e.g., see][]{F2016}. Second, we find an increase of the source, hybrid and $\beta$-dependent ({although} in this case it is less important) integral terms when the magnetic guide field increases. For Run III, the source and hybrid terms {become} even dominant (in absolute value) at small scales{;} however, since they have the same amplitude but with a different sign they cancel each other leaving the compressible flux as the main contribution to the cascade rate. Still for Run III, it is interesting to note that it is precisely when the compressible flux dominates (in absolute value) that the source and hybrid terms behave differently. Finally, we see that the compressible cascade rate $\varepsilon_C$ is more difficult to evaluate in {the} presence of $B_0$ because the inertial range becomes narrower (a higher spatial resolution seems to be necessary to get a reliable evaluation {of this quantity}). Note that in this case the fluctuating kinetic and magnetic energies become smaller (by a factor of ${\approx} 3$) in comparison with the cases without guide field{, resulting from the fact that we kept the forcing amplitude fixed for all simulations independently of the value of $B_0$.}

\begin{figure}
\begin{center}
	\includegraphics[width=.35\textwidth]{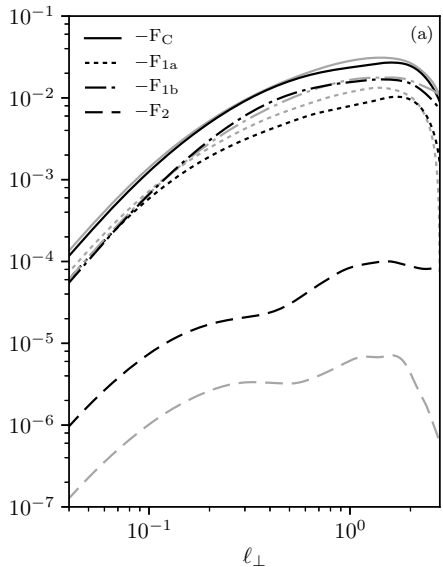}
\end{center}
\caption{(a) Total compressible flux F$_\text{C}$ (solid) and its components F$_{1a}$ (dashed-dot), F$_{1b}$ (dot) and F$_{2}$ (dashed) as a function of $\ell_\perp$, for Runs I (gray) and IV (black).}\label{fig4a}
\end{figure}

\begin{figure}
\begin{center}
	\includegraphics[width=.23\textwidth]{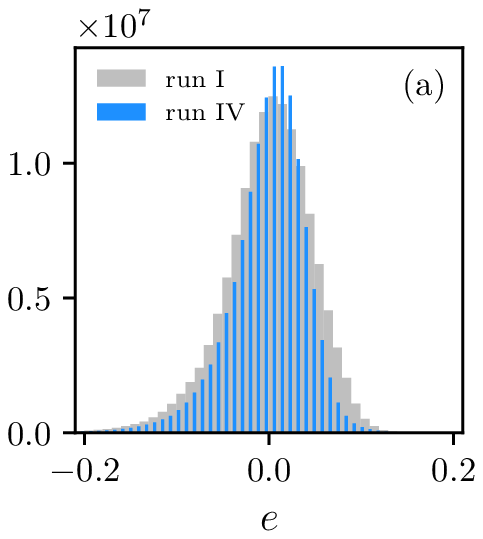}\includegraphics[width=.23\textwidth]{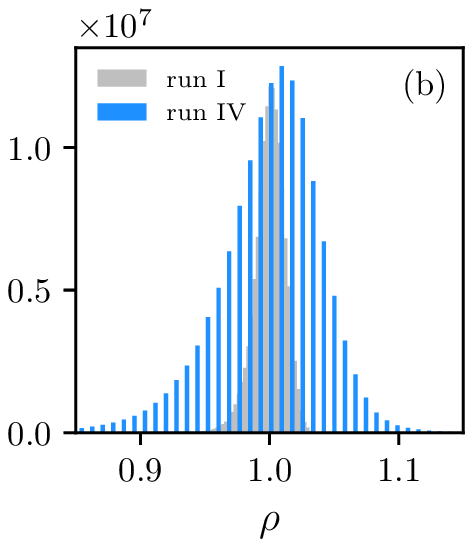} \\ \includegraphics[width=.23\textwidth]{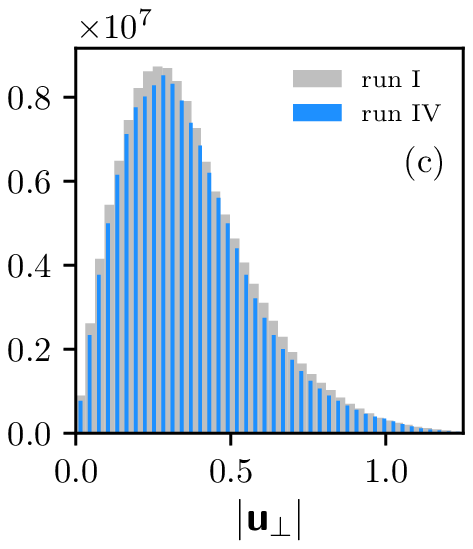}\includegraphics[width=.23\textwidth]{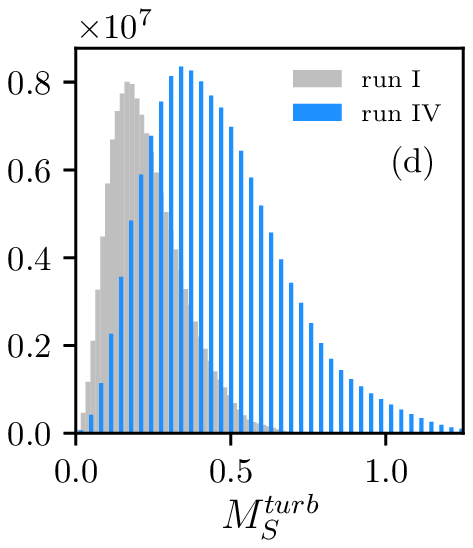}
\end{center}
\caption{Histograms of the internal energy $e$ (a), the mass density $\rho$ (b), the absolute value of the perpendicular velocity $| \textbf{u}_\perp|$ (c) and the turbulent sonic Mach number $M_S^{turb}$ (d), for runs I (grey) and IV (black lines).}\label{fig4b}
\end{figure}

%-------------------------------------------------------------
\subsection{Flux term}\label{results_flux}
%-------------------------------------------------------------

The compressible flux in {Eq.~}\eqref{term_flux} can be decomposed as $\textbf{F}_\text{C}=\textbf{F}_{1a}+\textbf{F}_{1b}+\textbf{F}_{2}$ with
\begin{align}
	\textbf{F}_{1a} \equiv &~\big\langle[(\delta(\rho\uh)\cdot\delta\uh+\delta(\rho\ua)\cdot\delta\ua\big]\delta\uh\rangle , \\
	\textbf{F}_{1b} \equiv &-\langle[\delta(\rho\uh)\cdot\delta\ua+\delta\uh\cdot\delta(\rho\ua)]\delta\ua\big\rangle,  \\
	\textbf{F}_2 \equiv & ~2\langle\delta e\delta\rho\delta\uh\rangle.
\end{align}
The term $\textbf{F}_\text{1} = \textbf{F}_{1a}+\textbf{F}_{1b}$ can be identified as the compressible version of the (incompressible) MHD Yaglom flux \citep{MY1975} and $\textbf{F}_\text{2}$ corresponds to a new purely compressible flux. Figure \ref{fig4a} shows the total compressible flux F$_\text{C}$ and its components F$_{1a}$, F$_{1b}$ and F$_{2}$ as a function of $\ell_\perp$ for Runs I and IV ({both with} $B_0=0$) for $M_S=0.25$ and $M_S=0.5$ respectively. {Figure} \ref{fig4b} displays the histograms over all the numerical {domain} of the internal compressible energy {density} $e$, {mass} density values $\rho$, the absolute value of the perpendicular velocity $|\textbf{u}_\perp|${,} and the {pointwise} turbulent Mach number $M_S^{turb}\equiv {u}/c_s$ for Runs I and IV. {In Fig.~\ref{fig4a} one can see{, in comparisson with Run I, that in run IV, which has a larger} Mach number{, the purely compressible component F$_2$ is also significantly larger} (at least one order of magnitude through all spatial perpendicular scales), while the Yaglom-like terms F$_{1a}$ and F$_{1b}$ remain approximately the same}. Furthermore, while $e$ and $|\textbf{u}_\perp|$ have almost the same statistical values for both runs, the distribution of density values for $M_S=0.5$ has a larger spread around the reference density value ($\rho_0=1$) than the one for $M_S=0.25$. {{Also}, we obtain a distribution for the internal energy {density $e$ which} is compatible with previous results in the literature \citep[e.g., see][]{PV1998,F2010,FB2015,N2015}. Note that the {statistical properties of the} internal energy is relevant for star formation dynamics \citep{FK2012}}. The large spread in {mass} density {fluctuations} plus the different turbulent Mach numbers in both runs explain the strong increase in {the} amplitude of F$_2$. However, we see that even for $M_S=0.5$ the contribution of F$_2$ to the total compressible flux remains negligible, which may be explained by the relatively low density fluctuations $\delta \rho/\rho \lesssim 10\%$ as can be seen in Fig.~\ref{fig4b} (b). Therefore, for small initial values of the sonic Mach number and zero magnetic guide field, {and for the range of parameters considered in this study, we conclude that} the dominant contribution {to} the total compressible flux is due to the Yaglom-like terms.

{Finally, we recall that in the present runs we used a solenoidal mechanical forcing for the velocity field. In runs with a balanced solenoidal/compressible external forcing one may expect to obtain different results. This issue is particularly relevant for distant astrophysical plasmas such as the interstellar medium or supernova remnants \citep{F2010,F2017}, where compressible forcing plays an important role in the injection of energy in the system.}

\begin{figure}
	\begin{center}
		\includegraphics[width=.4\textwidth]{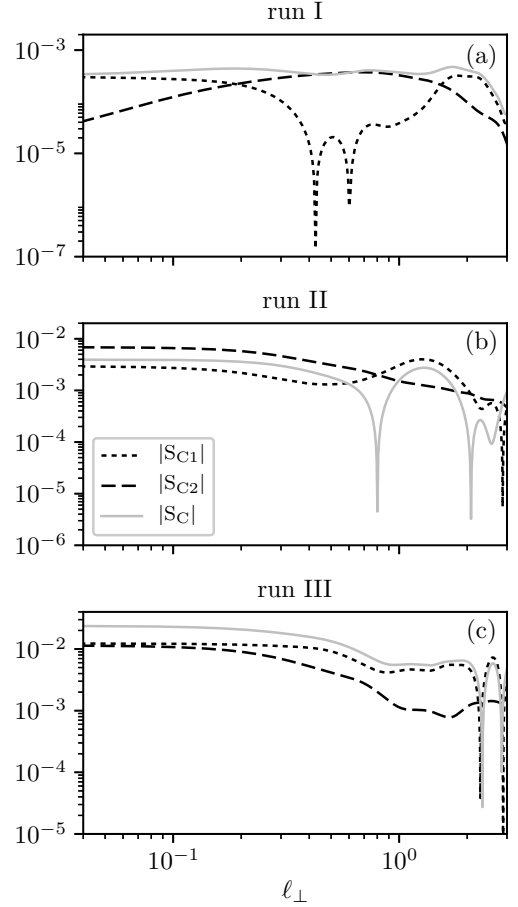}
	\end{center}
	\caption{Total source term S$_\text{C}$ (light gray) and its components S$_\text{C1}$ (dot black) and S$_\text{C2}$ (dashed black) as a function of $\ell_\perp$ for Runs I (a) , II (b) and III (c).}
	\label{fig5}
\end{figure}

\begin{figure}
	\begin{center}
		\includegraphics[width=.4\textwidth]{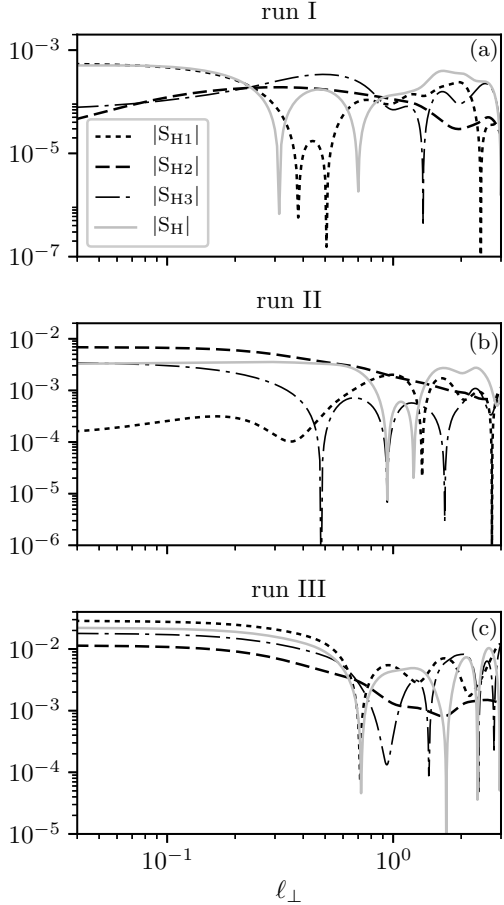}
	\end{center}
	\caption{Total hybrid term S$_\text{H}$ (light gray) and its components S$_\text{H1}$ (dot black), S$_{H2}$ (dashed black) and S$_\text{H3}$ (dashed-dot black) as a function of $\ell_\perp$ for Runs I (a), II (b) and III (c).}
	\label{fig6}
\end{figure}

\begin{figure}
	\begin{center}
		\includegraphics[width=.4\textwidth]{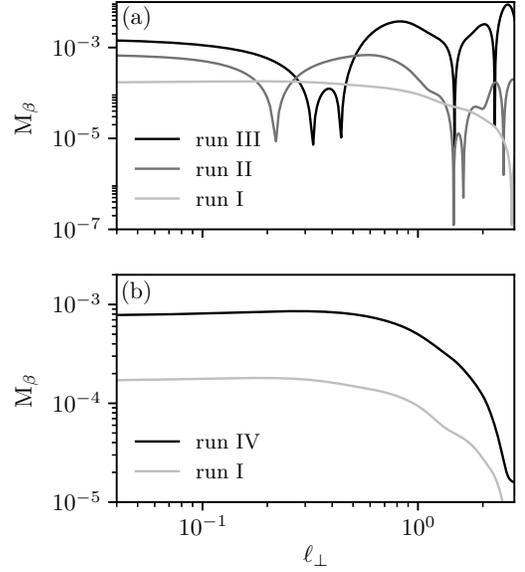}
	\end{center}
	\caption{$\beta$-dependent terms M$_\beta$ as a function of $\ell_\perp$ for (a) Runs I (light gray), II (gray) and III (black), and for (b) Run I (light gray) and IV (black), respectively.}
	\label{fig7}
\end{figure}

%-------------------------------------------------------------
\subsection{Source, hybrid and $\beta$-dependent terms}\label{results_source}
%-------------------------------------------------------------

The source, hybrid and $\beta$-dependent terms of the exact law \eqref{exactlaw} arise exclusively because of the compressibility of the plasma \citep{A2017a} (in the incompressible case they are exactly null). In particular, while the source and hybrid terms are proportional to $\dv$, $\da$ and $\boldsymbol{\nabla}\cdot(\rho\uh)$, the mixed $\beta$-dependent term is proportional to $\boldsymbol{\nabla}\cdot(e'\rho\uh)$. All these terms may modify the energy cascade rate in the inertial range, which is assumed to be constant at those scales.

The source \eqref{term_source} can be cast as $\text{S}_\text{C} =\text{S}_\text{C1} + \text{S}_\text{C2}$, with
\begin{align}\nonumber
	\text{S}_\text{C1}\equiv&~\langle[R_E'-\frac{1}{2}(R_B'+R_B)](\dv) \\
	&+[R_E-\frac{1}{2}(R_B+R_B')](\dvp)\rangle , \\\nonumber
	\text{S}_\text{C2} \equiv&~ \langle[(R_H-R_H')-\bar{\rho}(\uh'\cdot\ua)](\da) \\
	&+[(R_H'-R_H)-\bar{\rho}(\uh\cdot\ua')](\dap)\rangle,
\end{align}
{where} S$_\text{C1}$ and S$_\text{C2}$ correspond to the terms proportional to $\dv$ and $\da$, respectively. The hybrid term \eqref{term_hybrid} (which{, as already mentioned,} can be expressed as a source or flux-like {term} \citep{B2013,A2017b}) can be cast as $\text{S}_\text{H} =\text{S}_\text{H1} + \text{S}_\text{H2} + \text{S}_\text{H3}$, with
\begin{align}
	\text{S}_\text{H1} \equiv& \langle\big(\frac{P_M'-P'}{2}-E'\big)(\dv)+\big(\frac{P_M-P}{2}-E\big)(\dvp)\rangle , \\
	\text{S}_\text{H2} \equiv& \langle H'(\da)+H(\dap)\rangle , \\
	\text{S}_\text{H3} \equiv&\frac{1}{2}\langle\big(e'+\frac{u_{A}^{'2}}{2}\big)\big[\boldsymbol\nabla\cdot(\rho\uh)\big] +\big(e+\frac{u_{A}^{2}}{2}\big)\big[\boldsymbol\nabla'\cdot(\rho'\uh')\big]\rangle,
\end{align}
{where} S$_\text{H1}$, S$_\text{H2}$ and S$_\text{H3}$ correspond to the terms proportional to $\dv$, $\da$ and $\boldsymbol{\nabla}\cdot(\rho\uh)$, respectively. Note that in recent observational works \citep{B2016c,H2017a}, only the component S$_\text{H3}$ was used to compute the solar wind energy cascade rate, {besides the flux terms of Eq.~{(\ref{term_flux})}}. The rest of the hybrid components, i.e.{,} S$_\text{H1}$ and S$_\text{H2}$, were assumed to be sub-dominant in the inertial range. We will return to this point in Sec. \ref{discussion}.

{Figures} \ref{fig5} and \ref{fig6} show the absolute values of the source and hybrid terms as {a} function of $\ell_\perp$ for Runs I, II and III. Like above{, the} total (integrated) source and hybrid terms increase with increasing magnetic guide field (but {while} keeping the sonic Mach number constant). This behavior reflects the fact that S$_\text{C}$ and S$_\text{H}$ are explicitly proportional to $\textbf{B}_0$ since $\ua$ includes the mean plus the fluctuat{ing} magnetic field. Furthermore, both terms tend to the same value in the small-scale limit.

Under the assumption of statistical stationarity of the $\beta$ parameter, the $\beta$-dependent term \eqref{term_beta} can be converted into flux-like and be more easily {measured} using single-spacecraft data \citep[see,][]{B2016c,H2017a,A2017b}. However, in the present paper, we do not assume such additional hypothesis about the $\beta$ parameter. {Figure} \ref{fig7}(a) displays the total $\beta$-dependent term M$_\beta$ as a function of $\ell_\perp$ for $B_0=0$, $B_0=2$ and $B_0=8$ with $M_S=0.25$ (i.e., Runs I, II and III respectively) while Fig.~\ref{fig7}(b) shows the same quantity for $M_S=0.25$ and $M_S=0.5$ with $B_0=0$ (Runs I and IV respectively). As for the other contributions, when we increase the magnetic guide field, the $\beta$-dependent term increases. We see, however, that it remains mainly smaller than the other contributions and in particular smaller than the compressible flux, which is compatible with the analysis shown in Figs.~\ref{fig1} to \ref{fig3}. Finally, from Fig.~\ref{fig7}(b) we note that this term has a strong dependence on the Mach number, as does the mass density {fluctuations} (see Fig.~\ref{fig4b}). This can be also concluded directly from Eq. \eqref{term_beta}. {Note that i}n contrast to previous results \citep{H2017a}, here we consider the total density values, i.e., the mean plus the fluctuat{ing component.}

%-------------------------------------------------------------
\section{Discussion and Conclusion}\label{discussion}
%-------------------------------------------------------------

We presented {the} first detailed 3D numerical analysis of the exact law for fully developed isothermal CMHD turbulence \citep{A2017b,B2013}. Following \citet{A2017b}, we have separated the different contributions of the exact law in four types of terms, i.e.{,} the compressible flux, source, hybrid and $\beta$-dependent terms. We run different simulations with varying initial Mach number and magnetic guide field. For all the runs, the compressible flux {was} found to be the dominant component in the exact law for CMHD turbulence. Furthermore, and as expected, this term is not strongly affected by the presence of a magnetic guide field $\textbf{B}_0$ since it is a product of increments (and because the total density does not vary significantly between two points in space). In contrast, $\textbf{B}_0$ was found to have a strong impact on the remaining terms of the exact law \eqref{exactlaw} \citep[see also, ][]{A2017b} and also on the anisotropy of the {flow} \citep{Sh1983,M1996,Ma1999,Ga2000,Br2010,W2012,Ou2013,Me2016,Ou2016,S2016,F2016,A2017a}. Our numerical findings show a clear increase in $\text{S}_\text{C}$, $\text{S}_\text{H}$ and M$_\beta$ terms as $B_0$ is increased from 0 to 8. However, in all these cases the addition of these terms remain negligible with respect to the total compressible flux. Therefore, our energy cascade rate estimate has only a weak dependence on the magnetic guide field. {It is worth mentioning that this result may be quite different if we consider the case of a strong guide field ($B_0>10$), supersonic turbulence ($M_S>1$){,} and/or compressi{ble} driving of the velocity field.}

Using in-situ measurements from the THEMIS mission, \citet{B2016c} and \citet{H2017a} have investigated the role of compressible fluctuations in the MHD energy cascade rate for the fast and slow solar winds. {Those works were extended recently to the terrestrial magnetosheath where a first estimation of the energy cascade rate {was} obtained~\citep{H2017b}}. The authors computed some of the terms of the exact law \eqref{exactlaw} and compared their relative impact on the total compressible energy cascade rate $\varepsilon_C$. In these original works, the authors used an isotropic decomposition to compute the Yaglom-like term (i.e.,  $\textbf{F}_1$), the compressible flux (i.e., $\textbf{F}_2$) and a third flux-like term $\textbf{F}_3$, which is a {combination} of a part of the hybrid and the $\beta$-dependent (assuming statistical stationarity of $\beta$) terms. In particular,
{
\begin{align}\label{f3}
	\textbf{F}_3 = 2\left<\left(\bar{e}+\overline{\beta^{-1}e}+\frac{\overline{u_\text{A}^{2}}}{2}\right)\delta(\rho_1\uh)\right>,
\end{align}}
where $\rho_1$ corresponds to the density fluctuations (the part proportional to $\rho_0$ has been written as a source and has not been computed). It is straightforward to identify the parts of $\text{S}_\text{H3}$ and M$_\beta$ which are involved in Eq. \eqref{f3}.

In Refs. \citep{B2016c,H2017a}, the authors have found for the majority of the analyzed events comparable values of the compressible energy cascade rate $\varepsilon_C$ and the incompressible one $\varepsilon_I$ (computed from the exact law for IMHD turbulence \citep{P1998a,P1998b}). That statistical result is compatible with our numerical findings, in which the Yaglom-like flux is the dominant component of Eq. \eqref{exactlaw} and is very close to the incompressible Yaglom term \citep{MY1975}. However, some of the spacecraft observations showed that the compressible Yaglom flux and/or the ($\textbf{F}_2+\textbf{F}_3$) term can play a leading role in amplifying $\varepsilon_C$ with respect to $\varepsilon_I$, in particular in the slow solar wind (see Fig. 10 in~\citet{H2017a}). There are two possible explanations to those situations, which are not necessarily mutually exclusive. First, those events have larger density (and magnetic field) fluctuations that go beyond the values covered by our simulations in particular in the slow solar wind where $\delta \rho/\rho \lesssim 20\%$ and the turbulent Mach number ${M_S}^{turb}\lesssim 0.8$. This should be particularly true for the events that showed higher ratio of $\textbf{F}_1/\textbf{F}_I$ up to $10$ (see Fig. 10 in \citep{H2017a}). The other possibility is that some missing (source) terms would have compensated (at least partly) the $\textbf{F}_3$ term in those works, as we showed in the present simulations. Indeed, as recalled above, the observational results in Refs. \citep{B2016c,H2017a,H2017b} considered only the contributions from $\text{S}_\text{H3}$, while our simulations results indicate that the other terms, $\text{S}_\text{H1}$ and $\text{S}_\text{H2}$, may well have equal contribution, and consequently should be considered. As we mentioned in Sec.~\ref{results_source}, the compressible source terms involve local divergences that cannot be computed reliably using a single spacecraft because of the entanglement of the space and time variations (see Eq.~\ref{term_source}). Thus, in Refs. \citep{B2016c,H2017a,H2017b}, the authors had to assume that those terms are sub-dominant in the inertial range (this was also based on numerical simulations of supersonic hydrodynamic turbulence \citep{K2013}). A future improvement of those observational works would be to try to estimate the missing (source and hybrid) terms using multispacecraft data from the Cluster or the MMS mission to evaluate the local vector field divergences. However, such methods remain to be developed. From the numerical viewpoint, {an} improvement {to} the present work would consist in making the code capable of capturing higher density fluctuations and higher Mach numbers than those studied here. This is needed to meet the physical conditions observed in particular in planetary magnetosheaths~\citep{H2017b}. These problems will be investigated in forthcoming works.

\section*{Acknowledgments}

N.A. is supported through {a DIM-ACAV post-doctoral fellowship} and by LABEX Plas@Par through a grant managed by the Agence Nationale de la Recherche (ANR), as part of the program “Investissements d’Avenir” under the reference ANR-11-IDEX-0004–02. F.S., S.G. and N.A. acknowledge financial support from Programme National Soleil--Terre (PNST). P.D{.} and P.D.M. acknowledge support from UBACYT Grant No.~20020130100738BA and PICT  Grant {No.~2015-3530.} This work was granted access to the HPC resources of CINES under allocation 2017 A0030407714 made by GENCI. NA acknowledges Luis N. Martin for useful discussions.

%\bibliographystyle{apsrev4-1}
%\bibliography{cites}

\begin{thebibliography}{73}%
\makeatletter
\providecommand \@ifxundefined [1]{%
 \@ifx{#1\undefined}
}%
\providecommand \@ifnum [1]{%
 \ifnum #1\expandafter \@firstoftwo
 \else \expandafter \@secondoftwo
 \fi
}%
\providecommand \@ifx [1]{%
 \ifx #1\expandafter \@firstoftwo
 \else \expandafter \@secondoftwo
 \fi
}%
\providecommand \natexlab [1]{#1}%
\providecommand \enquote  [1]{``#1''}%
\providecommand \bibnamefont  [1]{#1}%
\providecommand \bibfnamefont [1]{#1}%
\providecommand \citenamefont [1]{#1}%
\providecommand \href@noop [0]{\@secondoftwo}%
\providecommand \href [0]{\begingroup \@sanitize@url \@href}%
\providecommand \@href[1]{\@@startlink{#1}\@@href}%
\providecommand \@@href[1]{\endgroup#1\@@endlink}%
\providecommand \@sanitize@url [0]{\catcode `\\12\catcode `\$12\catcode
  `\&12\catcode `\#12\catcode `\^12\catcode `\_12\catcode `\%12\relax}%
\providecommand \@@startlink[1]{}%
\providecommand \@@endlink[0]{}%
\providecommand \url  [0]{\begingroup\@sanitize@url \@url }%
\providecommand \@url [1]{\endgroup\@href {#1}{\urlprefix }}%
\providecommand \urlprefix  [0]{URL }%
\providecommand \Eprint [0]{\href }%
\providecommand \doibase [0]{http://dx.doi.org/}%
\providecommand \selectlanguage [0]{\@gobble}%
\providecommand \bibinfo  [0]{\@secondoftwo}%
\providecommand \bibfield  [0]{\@secondoftwo}%
\providecommand \translation [1]{[#1]}%
\providecommand \BibitemOpen [0]{}%
\providecommand \bibitemStop [0]{}%
\providecommand \bibitemNoStop [0]{.\EOS\space}%
\providecommand \EOS [0]{\spacefactor3000\relax}%
\providecommand \BibitemShut  [1]{\csname bibitem#1\endcsname}%
\let\auto@bib@innerbib\@empty
%</preamble>
\bibitem [{\citenamefont {Frisch}(1995)}]{F1995}%
  \BibitemOpen
  \bibfield  {author} {\bibinfo {author} {\bibfnamefont {U.}~\bibnamefont
  {Frisch}},\ }\href@noop {} {\emph {\bibinfo {title} {Turbulence: The Legacy
  of A. N. Kolmogorov}}}\ (\bibinfo  {publisher} {Cambridge University
  Press.},\ \bibinfo {year} {1995})\BibitemShut {NoStop}%
\bibitem [{\citenamefont {Kolmogorov}(1941{\natexlab{a}})}]{K1941a}%
  \BibitemOpen
  \bibfield  {author} {\bibinfo {author} {\bibfnamefont {A.~N.}\ \bibnamefont
  {Kolmogorov}},\ }in\ \href@noop {} {\emph {\bibinfo {booktitle} {Dokl. Akad.
  Nauk SSSR}}},\ Vol.~\bibinfo {volume} {30}\ (\bibinfo {year} {1941})\ pp.\
  \bibinfo {pages} {299--303}\BibitemShut {NoStop}%
\bibitem [{\citenamefont {Kolmogorov}(1941{\natexlab{b}})}]{K1941b}%
  \BibitemOpen
  \bibfield  {author} {\bibinfo {author} {\bibfnamefont {A.~N.}\ \bibnamefont
  {Kolmogorov}},\ }in\ \href@noop {} {\emph {\bibinfo {booktitle} {Dokl. Akad.
  Nauk SSSR}}},\ Vol.~\bibinfo {volume} {32}\ (\bibinfo {year} {1941})\ pp.\
  \bibinfo {pages} {16--18}\BibitemShut {NoStop}%
\bibitem [{\citenamefont {Chandrasekhar}(1951)}]{Ch1951}%
  \BibitemOpen
  \bibfield  {author} {\bibinfo {author} {\bibfnamefont {S.}~\bibnamefont
  {Chandrasekhar}},\ }\href@noop {} {\bibfield  {journal} {\bibinfo  {journal}
  {Proc. R. Soc. Lond. A}\ }\textbf {\bibinfo {volume} {204}},\ \bibinfo
  {pages} {435} (\bibinfo {year} {1951})}\BibitemShut {NoStop}%
\bibitem [{\citenamefont {Politano}\ and\ \citenamefont
  {Pouquet}(1998{\natexlab{a}})}]{P1998a}%
  \BibitemOpen
  \bibfield  {author} {\bibinfo {author} {\bibfnamefont {H.}~\bibnamefont
  {Politano}}\ and\ \bibinfo {author} {\bibfnamefont {A.}~\bibnamefont
  {Pouquet}},\ }\href@noop {} {\bibfield  {journal} {\bibinfo  {journal}
  {Physical Review E}\ }\textbf {\bibinfo {volume} {57}},\ \bibinfo {pages}
  {R21} (\bibinfo {year} {1998}{\natexlab{a}})}\BibitemShut {NoStop}%
\bibitem [{\citenamefont {Politano}\ and\ \citenamefont
  {Pouquet}(1998{\natexlab{b}})}]{P1998b}%
  \BibitemOpen
  \bibfield  {author} {\bibinfo {author} {\bibfnamefont {H.}~\bibnamefont
  {Politano}}\ and\ \bibinfo {author} {\bibfnamefont {A.}~\bibnamefont
  {Pouquet}},\ }\href@noop {} {\bibfield  {journal} {\bibinfo  {journal}
  {Geophysical Research Letters}\ }\textbf {\bibinfo {volume} {25}},\ \bibinfo
  {pages} {273} (\bibinfo {year} {1998}{\natexlab{b}})}\BibitemShut {NoStop}%
\bibitem [{\citenamefont {Mininni}\ and\ \citenamefont
  {Pouquet}(2009)}]{Mi2009}%
  \BibitemOpen
  \bibfield  {author} {\bibinfo {author} {\bibfnamefont {P.~D.}\ \bibnamefont
  {Mininni}}\ and\ \bibinfo {author} {\bibfnamefont {A.}~\bibnamefont
  {Pouquet}},\ }\href@noop {} {\bibfield  {journal} {\bibinfo  {journal} {Phys.
  Rev. E}\ }\textbf {\bibinfo {volume} {80}},\ \bibinfo {pages} {025401}
  (\bibinfo {year} {2009})}\BibitemShut {NoStop}%
\bibitem [{\citenamefont {{Bhattacharjee}}\ \emph {et~al.}(2009)\citenamefont
  {{Bhattacharjee}}, \citenamefont {{Huang}}, \citenamefont {{Yang}},\ and\
  \citenamefont {{Rogers}}}]{B2009}%
  \BibitemOpen
  \bibfield  {author} {\bibinfo {author} {\bibfnamefont {A.}~\bibnamefont
  {{Bhattacharjee}}}, \bibinfo {author} {\bibfnamefont {Y.-M.}\ \bibnamefont
  {{Huang}}}, \bibinfo {author} {\bibfnamefont {H.}~\bibnamefont {{Yang}}}, \
  and\ \bibinfo {author} {\bibfnamefont {B.}~\bibnamefont {{Rogers}}},\
  }\href@noop {} {\bibfield  {journal} {\bibinfo  {journal} {Phys. Plasmas}\
  }\textbf {\bibinfo {volume} {16}},\ \bibinfo {pages} {112102} (\bibinfo
  {year} {2009})}\BibitemShut {NoStop}%
\bibitem [{\citenamefont {Wan}\ \emph {et~al.}(2010)\citenamefont {Wan},
  \citenamefont {Servidio}, \citenamefont {Oughton},\ and\ \citenamefont
  {Matthaeus}}]{W2010}%
  \BibitemOpen
  \bibfield  {author} {\bibinfo {author} {\bibfnamefont {M.}~\bibnamefont
  {Wan}}, \bibinfo {author} {\bibfnamefont {S.}~\bibnamefont {Servidio}},
  \bibinfo {author} {\bibfnamefont {S.}~\bibnamefont {Oughton}}, \ and\
  \bibinfo {author} {\bibfnamefont {W.~H.}\ \bibnamefont {Matthaeus}},\
  }\href@noop {} {\bibfield  {journal} {\bibinfo  {journal} {Phys. Plasmas}\
  }\textbf {\bibinfo {volume} {17}},\ \bibinfo {pages} {052307} (\bibinfo
  {year} {2010})}\BibitemShut {NoStop}%
\bibitem [{\citenamefont {Yoshimatsu}(2012)}]{Y2012}%
  \BibitemOpen
  \bibfield  {author} {\bibinfo {author} {\bibfnamefont {K.}~\bibnamefont
  {Yoshimatsu}},\ }\href@noop {} {\bibfield  {journal} {\bibinfo  {journal}
  {Phys. Rev. E}\ }\textbf {\bibinfo {volume} {85}},\ \bibinfo {pages} {066313}
  (\bibinfo {year} {2012})}\BibitemShut {NoStop}%
\bibitem [{\citenamefont {Andr{\'e}s}\ \emph
  {et~al.}(2016{\natexlab{a}})\citenamefont {Andr{\'e}s}, \citenamefont
  {Mininni}, \citenamefont {Dmitruk},\ and\ \citenamefont {Gomez}}]{A2016b}%
  \BibitemOpen
  \bibfield  {author} {\bibinfo {author} {\bibfnamefont {N.}~\bibnamefont
  {Andr{\'e}s}}, \bibinfo {author} {\bibfnamefont {P.~D.}\ \bibnamefont
  {Mininni}}, \bibinfo {author} {\bibfnamefont {P.}~\bibnamefont {Dmitruk}}, \
  and\ \bibinfo {author} {\bibfnamefont {D.~O.}\ \bibnamefont {Gomez}},\
  }\href@noop {} {\bibfield  {journal} {\bibinfo  {journal} {Physical Review
  E}\ }\textbf {\bibinfo {volume} {93}},\ \bibinfo {pages} {063202} (\bibinfo
  {year} {2016}{\natexlab{a}})}\BibitemShut {NoStop}%
\bibitem [{\citenamefont {Sorriso-Valvo}\ \emph {et~al.}(2007)\citenamefont
  {Sorriso-Valvo}, \citenamefont {Marino}, \citenamefont {Carbone},
  \citenamefont {Noullez}, \citenamefont {Lepreti}, \citenamefont {Veltri},
  \citenamefont {Bruno}, \citenamefont {Bavassano},\ and\ \citenamefont
  {Pietropaolo}}]{SV2007}%
  \BibitemOpen
  \bibfield  {author} {\bibinfo {author} {\bibfnamefont {L.}~\bibnamefont
  {Sorriso-Valvo}}, \bibinfo {author} {\bibfnamefont {R.}~\bibnamefont
  {Marino}}, \bibinfo {author} {\bibfnamefont {V.}~\bibnamefont {Carbone}},
  \bibinfo {author} {\bibfnamefont {A.}~\bibnamefont {Noullez}}, \bibinfo
  {author} {\bibfnamefont {F.}~\bibnamefont {Lepreti}}, \bibinfo {author}
  {\bibfnamefont {P.}~\bibnamefont {Veltri}}, \bibinfo {author} {\bibfnamefont
  {R.}~\bibnamefont {Bruno}}, \bibinfo {author} {\bibfnamefont
  {B.}~\bibnamefont {Bavassano}}, \ and\ \bibinfo {author} {\bibfnamefont
  {E.}~\bibnamefont {Pietropaolo}},\ }\href@noop {} {\bibfield  {journal}
  {\bibinfo  {journal} {Physical review letters}\ }\textbf {\bibinfo {volume}
  {99}},\ \bibinfo {pages} {115001} (\bibinfo {year} {2007})}\BibitemShut
  {NoStop}%
\bibitem [{\citenamefont {Weygand}\ \emph {et~al.}(2007)\citenamefont
  {Weygand}, \citenamefont {Matthaeus}, \citenamefont {Dasso}, \citenamefont
  {Kivelson},\ and\ \citenamefont {Walker}}]{WEY2007}%
  \BibitemOpen
  \bibfield  {author} {\bibinfo {author} {\bibfnamefont {J.~M.}\ \bibnamefont
  {Weygand}}, \bibinfo {author} {\bibfnamefont {W.~H.}\ \bibnamefont
  {Matthaeus}}, \bibinfo {author} {\bibfnamefont {S.}~\bibnamefont {Dasso}},
  \bibinfo {author} {\bibfnamefont {M.~G.}\ \bibnamefont {Kivelson}}, \ and\
  \bibinfo {author} {\bibfnamefont {R.~J.}\ \bibnamefont {Walker}},\
  }\href@noop {} {\bibfield  {journal} {\bibinfo  {journal} {J. Geophys. Res.:
  Space Phys.}\ }\textbf {\bibinfo {volume} {112}},\ \bibinfo {pages} {A10}
  (\bibinfo {year} {2007})}\BibitemShut {NoStop}%
\bibitem [{\citenamefont {Marino}\ \emph {et~al.}(2008)\citenamefont {Marino},
  \citenamefont {Sorriso-Valvo}, \citenamefont {Carbone}, \citenamefont
  {Noullez}, \citenamefont {Bruno},\ and\ \citenamefont {Bavassano}}]{M2008}%
  \BibitemOpen
  \bibfield  {author} {\bibinfo {author} {\bibfnamefont {R.}~\bibnamefont
  {Marino}}, \bibinfo {author} {\bibfnamefont {L.}~\bibnamefont
  {Sorriso-Valvo}}, \bibinfo {author} {\bibfnamefont {V.}~\bibnamefont
  {Carbone}}, \bibinfo {author} {\bibfnamefont {A.}~\bibnamefont {Noullez}},
  \bibinfo {author} {\bibfnamefont {R.}~\bibnamefont {Bruno}}, \ and\ \bibinfo
  {author} {\bibfnamefont {B.}~\bibnamefont {Bavassano}},\ }\href@noop {}
  {\bibfield  {journal} {\bibinfo  {journal} {Astrophys. J. Lett.}\ }\textbf
  {\bibinfo {volume} {677}} (\bibinfo {year} {2008})}\BibitemShut {NoStop}%
\bibitem [{\citenamefont {Sahraoui}(2008)}]{Sa2008}%
  \BibitemOpen
  \bibfield  {author} {\bibinfo {author} {\bibfnamefont {F.}~\bibnamefont
  {Sahraoui}},\ }\href@noop {} {\bibfield  {journal} {\bibinfo  {journal}
  {Phys. Rev. E}\ }\textbf {\bibinfo {volume} {78}},\ \bibinfo {pages} {026402}
  (\bibinfo {year} {2008})}\BibitemShut {NoStop}%
\bibitem [{\citenamefont {Banerjee}\ and\ \citenamefont
  {Galtier}(2013)}]{B2013}%
  \BibitemOpen
  \bibfield  {author} {\bibinfo {author} {\bibfnamefont {S.}~\bibnamefont
  {Banerjee}}\ and\ \bibinfo {author} {\bibfnamefont {S.}~\bibnamefont
  {Galtier}},\ }\href@noop {} {\bibfield  {journal} {\bibinfo  {journal}
  {Physical Review E}\ }\textbf {\bibinfo {volume} {87}},\ \bibinfo {pages}
  {013019} (\bibinfo {year} {2013})}\BibitemShut {NoStop}%
\bibitem [{\citenamefont {Galtier}\ and\ \citenamefont
  {Banerjee}(2011)}]{Ga2011}%
  \BibitemOpen
  \bibfield  {author} {\bibinfo {author} {\bibfnamefont {S.}~\bibnamefont
  {Galtier}}\ and\ \bibinfo {author} {\bibfnamefont {S.}~\bibnamefont
  {Banerjee}},\ }\href@noop {} {\bibfield  {journal} {\bibinfo  {journal}
  {Physical review letters}\ }\textbf {\bibinfo {volume} {107}},\ \bibinfo
  {pages} {134501} (\bibinfo {year} {2011})}\BibitemShut {NoStop}%
\bibitem [{\citenamefont {Monin}\ and\ \citenamefont {Yaglom}(1975)}]{MY1975}%
  \BibitemOpen
  \bibfield  {author} {\bibinfo {author} {\bibfnamefont {A.~S.}\ \bibnamefont
  {Monin}}\ and\ \bibinfo {author} {\bibfnamefont {A.~M.}\ \bibnamefont
  {Yaglom}},\ }\href@noop {} {\emph {\bibinfo {title} {Statistical Fluid
  Mechanics: Mechanics of Turbulence}}},\ Vol.~\bibinfo {volume} {2}\ (\bibinfo
   {publisher} {Cambridge, MA: MIT Press.},\ \bibinfo {year}
  {1975})\BibitemShut {NoStop}%
\bibitem [{\citenamefont {Banerjee}\ and\ \citenamefont
  {Kritsuk}(2018{\natexlab{a}})}]{BF2018}%
  \BibitemOpen
  \bibfield  {author} {\bibinfo {author} {\bibfnamefont {S.}~\bibnamefont
  {Banerjee}}\ and\ \bibinfo {author} {\bibfnamefont {A.~G.}\ \bibnamefont
  {Kritsuk}},\ }\href@noop {} {\bibfield  {journal} {\bibinfo  {journal}
  {Physical Review E}\ }\textbf {\bibinfo {volume} {97}},\ \bibinfo {pages}
  {023107} (\bibinfo {year} {2018}{\natexlab{a}})}\BibitemShut {NoStop}%
\bibitem [{\citenamefont {Andr{\'e}s}\ and\ \citenamefont
  {Sahraoui}(2017)}]{A2017b}%
  \BibitemOpen
  \bibfield  {author} {\bibinfo {author} {\bibfnamefont {N.}~\bibnamefont
  {Andr{\'e}s}}\ and\ \bibinfo {author} {\bibfnamefont {F.}~\bibnamefont
  {Sahraoui}},\ }\href@noop {} {\bibfield  {journal} {\bibinfo  {journal}
  {Physical Review E}\ }\textbf {\bibinfo {volume} {96}},\ \bibinfo {pages}
  {053205} (\bibinfo {year} {2017})}\BibitemShut {NoStop}%
\bibitem [{\citenamefont {Carbone}\ \emph {et~al.}(2009)\citenamefont
  {Carbone}, \citenamefont {Marino}, \citenamefont {Sorriso-Valvo},
  \citenamefont {Noullez},\ and\ \citenamefont {Bruno}}]{C2009b}%
  \BibitemOpen
  \bibfield  {author} {\bibinfo {author} {\bibfnamefont {V.}~\bibnamefont
  {Carbone}}, \bibinfo {author} {\bibfnamefont {R.}~\bibnamefont {Marino}},
  \bibinfo {author} {\bibfnamefont {L.}~\bibnamefont {Sorriso-Valvo}}, \bibinfo
  {author} {\bibfnamefont {A.}~\bibnamefont {Noullez}}, \ and\ \bibinfo
  {author} {\bibfnamefont {R.}~\bibnamefont {Bruno}},\ }\href@noop {}
  {\bibfield  {journal} {\bibinfo  {journal} {Physical review letters}\
  }\textbf {\bibinfo {volume} {103}},\ \bibinfo {pages} {061102} (\bibinfo
  {year} {2009})}\BibitemShut {NoStop}%
\bibitem [{\citenamefont {Hadid}\ \emph {et~al.}(2017)\citenamefont {Hadid},
  \citenamefont {Sahraoui},\ and\ \citenamefont {Galtier}}]{H2017a}%
  \BibitemOpen
  \bibfield  {author} {\bibinfo {author} {\bibfnamefont {L.}~\bibnamefont
  {Hadid}}, \bibinfo {author} {\bibfnamefont {F.}~\bibnamefont {Sahraoui}}, \
  and\ \bibinfo {author} {\bibfnamefont {S.}~\bibnamefont {Galtier}},\
  }\href@noop {} {\bibfield  {journal} {\bibinfo  {journal} {The Astrophysical
  Journal}\ }\textbf {\bibinfo {volume} {838}},\ \bibinfo {pages} {9} (\bibinfo
  {year} {2017})}\BibitemShut {NoStop}%
\bibitem [{\citenamefont {Banerjee}\ \emph {et~al.}(2016)\citenamefont
  {Banerjee}, \citenamefont {Hadid}, \citenamefont {Sahraoui},\ and\
  \citenamefont {Galtier}}]{B2016c}%
  \BibitemOpen
  \bibfield  {author} {\bibinfo {author} {\bibfnamefont {S.}~\bibnamefont
  {Banerjee}}, \bibinfo {author} {\bibfnamefont {L.~Z.}\ \bibnamefont {Hadid}},
  \bibinfo {author} {\bibfnamefont {F.}~\bibnamefont {Sahraoui}}, \ and\
  \bibinfo {author} {\bibfnamefont {S.}~\bibnamefont {Galtier}},\ }\href@noop
  {} {\bibfield  {journal} {\bibinfo  {journal} {The Astrophysical Journal
  Letters}\ }\textbf {\bibinfo {volume} {829}},\ \bibinfo {pages} {L27}
  (\bibinfo {year} {2016})}\BibitemShut {NoStop}%
\bibitem [{\citenamefont {Hadid}\ \emph {et~al.}(2018)\citenamefont {Hadid},
  \citenamefont {Sahraoui}, \citenamefont {Galtier},\ and\ \citenamefont
  {Huang}}]{H2017b}%
  \BibitemOpen
  \bibfield  {author} {\bibinfo {author} {\bibfnamefont {L.}~\bibnamefont
  {Hadid}}, \bibinfo {author} {\bibfnamefont {F.}~\bibnamefont {Sahraoui}},
  \bibinfo {author} {\bibfnamefont {S.}~\bibnamefont {Galtier}}, \ and\
  \bibinfo {author} {\bibfnamefont {S.}~\bibnamefont {Huang}},\ }\href@noop {}
  {\bibfield  {journal} {\bibinfo  {journal} {Phys. Rev. Lett.}\ }\textbf
  {\bibinfo {volume} {120}},\ \bibinfo {pages} {055102} (\bibinfo {year}
  {2018})}\BibitemShut {NoStop}%
\bibitem [{\citenamefont {Zank}\ \emph {et~al.}(017a)\citenamefont {Zank},
  \citenamefont {Adhikari}, \citenamefont {Hunana}, \citenamefont {Shiota},
  \citenamefont {Bruno},\ and\ \citenamefont {Telloni}}]{Z2017a}%
  \BibitemOpen
  \bibfield  {author} {\bibinfo {author} {\bibfnamefont {G.}~\bibnamefont
  {Zank}}, \bibinfo {author} {\bibfnamefont {L.}~\bibnamefont {Adhikari}},
  \bibinfo {author} {\bibfnamefont {P.}~\bibnamefont {Hunana}}, \bibinfo
  {author} {\bibfnamefont {D.}~\bibnamefont {Shiota}}, \bibinfo {author}
  {\bibfnamefont {R.}~\bibnamefont {Bruno}}, \ and\ \bibinfo {author}
  {\bibfnamefont {D.}~\bibnamefont {Telloni}},\ }\href@noop {} {\bibfield
  {journal} {\bibinfo  {journal} {The Astrophysical Journal}\ }\textbf
  {\bibinfo {volume} {835}},\ \bibinfo {pages} {147} (\bibinfo {year}
  {2017a})}\BibitemShut {NoStop}%
\bibitem [{\citenamefont {Zank}\ and\ \citenamefont {Matthaeus}(1990)}]{Z1990}%
  \BibitemOpen
  \bibfield  {author} {\bibinfo {author} {\bibfnamefont {G.~P.}\ \bibnamefont
  {Zank}}\ and\ \bibinfo {author} {\bibfnamefont {W.~H.}\ \bibnamefont
  {Matthaeus}},\ }\href@noop {} {\bibfield  {journal} {\bibinfo  {journal}
  {Physical review letters}\ }\textbf {\bibinfo {volume} {64}},\ \bibinfo
  {pages} {1243} (\bibinfo {year} {1990})}\BibitemShut {NoStop}%
\bibitem [{\citenamefont {Zank}\ \emph {et~al.}(017b)\citenamefont {Zank},
  \citenamefont {Du},\ and\ \citenamefont {Hunana}}]{Z2017b}%
  \BibitemOpen
  \bibfield  {author} {\bibinfo {author} {\bibfnamefont {G.~P.}\ \bibnamefont
  {Zank}}, \bibinfo {author} {\bibfnamefont {S.}~\bibnamefont {Du}}, \ and\
  \bibinfo {author} {\bibfnamefont {P.}~\bibnamefont {Hunana}},\ }\href@noop {}
  {\bibfield  {journal} {\bibinfo  {journal} {The Astrophysical Journal}\
  }\textbf {\bibinfo {volume} {842}},\ \bibinfo {pages} {114} (\bibinfo {year}
  {2017b})}\BibitemShut {NoStop}%
\bibitem [{\citenamefont {Yang}\ \emph {et~al.}(2017)\citenamefont {Yang},
  \citenamefont {Matthaeus}, \citenamefont {Parashar}, \citenamefont
  {Haggerty}, \citenamefont {Roytershteyn}, \citenamefont {Daughton},
  \citenamefont {Wan}, \citenamefont {Shi},\ and\ \citenamefont
  {Chen}}]{Y2017}%
  \BibitemOpen
  \bibfield  {author} {\bibinfo {author} {\bibfnamefont {Y.}~\bibnamefont
  {Yang}}, \bibinfo {author} {\bibfnamefont {W.~H.}\ \bibnamefont {Matthaeus}},
  \bibinfo {author} {\bibfnamefont {T.~N.}\ \bibnamefont {Parashar}}, \bibinfo
  {author} {\bibfnamefont {C.~C.}\ \bibnamefont {Haggerty}}, \bibinfo {author}
  {\bibfnamefont {V.}~\bibnamefont {Roytershteyn}}, \bibinfo {author}
  {\bibfnamefont {W.}~\bibnamefont {Daughton}}, \bibinfo {author}
  {\bibfnamefont {M.}~\bibnamefont {Wan}}, \bibinfo {author} {\bibfnamefont
  {Y.}~\bibnamefont {Shi}}, \ and\ \bibinfo {author} {\bibfnamefont
  {S.}~\bibnamefont {Chen}},\ }\href@noop {} {\bibfield  {journal} {\bibinfo
  {journal} {Physics of Plasmas}\ }\textbf {\bibinfo {volume} {24}},\ \bibinfo
  {pages} {072306} (\bibinfo {year} {2017})}\BibitemShut {NoStop}%
\bibitem [{\citenamefont {{Grete}}\ \emph {et~al.}(2017)\citenamefont
  {{Grete}}, \citenamefont {{O'Shea}}, \citenamefont {{Beckwith}},
  \citenamefont {{Schmidt}},\ and\ \citenamefont {{Christlieb}}}]{Grete2017}%
  \BibitemOpen
  \bibfield  {author} {\bibinfo {author} {\bibfnamefont {P.}~\bibnamefont
  {{Grete}}}, \bibinfo {author} {\bibfnamefont {B.~W.}\ \bibnamefont
  {{O'Shea}}}, \bibinfo {author} {\bibfnamefont {K.}~\bibnamefont
  {{Beckwith}}}, \bibinfo {author} {\bibfnamefont {W.}~\bibnamefont
  {{Schmidt}}}, \ and\ \bibinfo {author} {\bibfnamefont {A.}~\bibnamefont
  {{Christlieb}}},\ }\href@noop {} {\bibfield  {journal} {\bibinfo  {journal}
  {Physics of Plasmas}\ }\textbf {\bibinfo {volume} {24}},\ \bibinfo {pages}
  {092311} (\bibinfo {year} {2017})}\BibitemShut {NoStop}%
\bibitem [{\citenamefont {Andr{\'e}s}\ \emph {et~al.}(2017)\citenamefont
  {Andr{\'e}s}, \citenamefont {Clark~di Leoni}, \citenamefont {Mininni},
  \citenamefont {Dmitruk}, \citenamefont {Sahraoui},\ and\ \citenamefont
  {Matthaeus}}]{A2017a}%
  \BibitemOpen
  \bibfield  {author} {\bibinfo {author} {\bibfnamefont {N.}~\bibnamefont
  {Andr{\'e}s}}, \bibinfo {author} {\bibfnamefont {P.}~\bibnamefont {Clark~di
  Leoni}}, \bibinfo {author} {\bibfnamefont {P.~D.}\ \bibnamefont {Mininni}},
  \bibinfo {author} {\bibfnamefont {P.}~\bibnamefont {Dmitruk}}, \bibinfo
  {author} {\bibfnamefont {F.}~\bibnamefont {Sahraoui}}, \ and\ \bibinfo
  {author} {\bibfnamefont {W.~H.}\ \bibnamefont {Matthaeus}},\ }\href@noop {}
  {\bibfield  {journal} {\bibinfo  {journal} {Physics of Plasmas}\ }\textbf
  {\bibinfo {volume} {24}},\ \bibinfo {pages} {102314} (\bibinfo {year}
  {2017})}\BibitemShut {NoStop}%
\bibitem [{\citenamefont {Andr{\'e}s}\ \emph {et~al.}(2018)\citenamefont
  {Andr{\'e}s}, \citenamefont {Galtier},\ and\ \citenamefont
  {Sahraoui}}]{A2018a}%
  \BibitemOpen
  \bibfield  {author} {\bibinfo {author} {\bibfnamefont {N.}~\bibnamefont
  {Andr{\'e}s}}, \bibinfo {author} {\bibfnamefont {S.}~\bibnamefont {Galtier}},
  \ and\ \bibinfo {author} {\bibfnamefont {F.}~\bibnamefont {Sahraoui}},\
  }\href@noop {} {\bibfield  {journal} {\bibinfo  {journal} {Physical Review
  E}\ }\textbf {\bibinfo {volume} {97}},\ \bibinfo {pages} {013204} (\bibinfo
  {year} {2018})}\BibitemShut {NoStop}%
\bibitem [{\citenamefont {Fitzpatrick}(2014)}]{F2014}%
  \BibitemOpen
  \bibfield  {author} {\bibinfo {author} {\bibfnamefont {R.}~\bibnamefont
  {Fitzpatrick}},\ }\href@noop {} {\emph {\bibinfo {title} {Plasma Physics: An
  Introduction}}}\ (\bibinfo  {publisher} {CRC Press},\ \bibinfo {year}
  {2014})\BibitemShut {NoStop}%
\bibitem [{\citenamefont {Marsch}\ and\ \citenamefont
  {Mangeney}(1987)}]{M1987}%
  \BibitemOpen
  \bibfield  {author} {\bibinfo {author} {\bibfnamefont {E.}~\bibnamefont
  {Marsch}}\ and\ \bibinfo {author} {\bibfnamefont {A.}~\bibnamefont
  {Mangeney}},\ }\href {\doibase 10.1029/JA092iA07p07363} {\bibfield  {journal}
  {\bibinfo  {journal} {Journal of Geophysical Research: Space Physics}\
  }\textbf {\bibinfo {volume} {92}},\ \bibinfo {pages} {7363} (\bibinfo {year}
  {1987})}\BibitemShut {NoStop}%
\bibitem [{\citenamefont {Duchon}\ and\ \citenamefont {Robert}(2000)}]{DR2000}%
  \BibitemOpen
  \bibfield  {author} {\bibinfo {author} {\bibfnamefont {J.}~\bibnamefont
  {Duchon}}\ and\ \bibinfo {author} {\bibfnamefont {R.}~\bibnamefont
  {Robert}},\ }\href {http://stacks.iop.org/0951-7715/13/i=1/a=312} {\bibfield
  {journal} {\bibinfo  {journal} {Nonlinearity}\ }\textbf {\bibinfo {volume}
  {13}},\ \bibinfo {pages} {249} (\bibinfo {year} {2000})}\BibitemShut
  {NoStop}%
\bibitem [{\citenamefont {Eyink}\ and\ \citenamefont {Drivas}(2018)}]{E2018}%
  \BibitemOpen
  \bibfield  {author} {\bibinfo {author} {\bibfnamefont {G.~L.}\ \bibnamefont
  {Eyink}}\ and\ \bibinfo {author} {\bibfnamefont {T.~D.}\ \bibnamefont
  {Drivas}},\ }\href {\doibase 10.1103/PhysRevX.8.011022} {\bibfield  {journal}
  {\bibinfo  {journal} {Phys. Rev. X}\ }\textbf {\bibinfo {volume} {8}},\
  \bibinfo {pages} {011022} (\bibinfo {year} {2018})}\BibitemShut {NoStop}%
\bibitem [{\citenamefont {Galtier}(2018)}]{Ga2018}%
  \BibitemOpen
  \bibfield  {author} {\bibinfo {author} {\bibfnamefont {S.}~\bibnamefont
  {Galtier}},\ }\href {http://stacks.iop.org/1751-8121/51/i=20/a=205501}
  {\bibfield  {journal} {\bibinfo  {journal} {Journal of Physics A:
  Mathematical and Theoretical}\ }\textbf {\bibinfo {volume} {51}},\ \bibinfo
  {pages} {205501} (\bibinfo {year} {2018})}\BibitemShut {NoStop}%
\bibitem [{\citenamefont {Andr{\'e}s}\ \emph
  {et~al.}(2016{\natexlab{b}})\citenamefont {Andr{\'e}s}, \citenamefont
  {Galtier},\ and\ \citenamefont {Sahraoui}}]{A2016c}%
  \BibitemOpen
  \bibfield  {author} {\bibinfo {author} {\bibfnamefont {N.}~\bibnamefont
  {Andr{\'e}s}}, \bibinfo {author} {\bibfnamefont {S.}~\bibnamefont {Galtier}},
  \ and\ \bibinfo {author} {\bibfnamefont {F.}~\bibnamefont {Sahraoui}},\
  }\href@noop {} {\bibfield  {journal} {\bibinfo  {journal} {Physical Review
  E}\ }\textbf {\bibinfo {volume} {94}},\ \bibinfo {pages} {063206} (\bibinfo
  {year} {2016}{\natexlab{b}})}\BibitemShut {NoStop}%
\bibitem [{\citenamefont {Banerjee}\ and\ \citenamefont
  {Kritsuk}(2018{\natexlab{b}})}]{B2018}%
  \BibitemOpen
  \bibfield  {author} {\bibinfo {author} {\bibfnamefont {S.}~\bibnamefont
  {Banerjee}}\ and\ \bibinfo {author} {\bibfnamefont {A.~G.}\ \bibnamefont
  {Kritsuk}},\ }\href {\doibase 10.1103/PhysRevE.97.023107} {\bibfield
  {journal} {\bibinfo  {journal} {Phys. Rev. E}\ }\textbf {\bibinfo {volume}
  {97}},\ \bibinfo {pages} {023107} (\bibinfo {year}
  {2018}{\natexlab{b}})}\BibitemShut {NoStop}%
\bibitem [{\citenamefont {Batchelor}(1953)}]{Ba1953}%
  \BibitemOpen
  \bibfield  {author} {\bibinfo {author} {\bibfnamefont {G.~K.}\ \bibnamefont
  {Batchelor}},\ }\href@noop {} {\emph {\bibinfo {title} {The theory of
  homogeneus turbulence}}}\ (\bibinfo  {publisher} {Cambridge Univ. Press},\
  \bibinfo {year} {1953})\BibitemShut {NoStop}%
\bibitem [{\citenamefont {de~K\'arm\'an}\ and\ \citenamefont
  {Howarth}(1938)}]{vkh1938}%
  \BibitemOpen
  \bibfield  {author} {\bibinfo {author} {\bibfnamefont {T.}~\bibnamefont
  {de~K\'arm\'an}}\ and\ \bibinfo {author} {\bibfnamefont {L.}~\bibnamefont
  {Howarth}},\ }\href {\doibase 10.1098/rspa.1938.0013} {\bibfield  {journal}
  {\bibinfo  {journal} {Proceedings of the Royal Society of London A:
  Mathematical, Physical and Engineering Sciences}\ }\textbf {\bibinfo {volume}
  {164}},\ \bibinfo {pages} {192} (\bibinfo {year} {1938})},\ \Eprint
  {http://arxiv.org/abs/http://rspa.royalsocietypublishing.org/content/164/917/192.full.pdf}
  {http://rspa.royalsocietypublishing.org/content/164/917/192.full.pdf}
  \BibitemShut {NoStop}%
\bibitem [{\citenamefont {Banerjee}\ and\ \citenamefont
  {Galtier}(2014)}]{B2014}%
  \BibitemOpen
  \bibfield  {author} {\bibinfo {author} {\bibfnamefont {S.}~\bibnamefont
  {Banerjee}}\ and\ \bibinfo {author} {\bibfnamefont {S.}~\bibnamefont
  {Galtier}},\ }\href@noop {} {\bibfield  {journal} {\bibinfo  {journal}
  {Journal of Fluid Mechanics}\ }\textbf {\bibinfo {volume} {742}},\ \bibinfo
  {pages} {230} (\bibinfo {year} {2014})}\BibitemShut {NoStop}%
\bibitem [{\citenamefont {Kritsuk}\ \emph {et~al.}(2013)\citenamefont
  {Kritsuk}, \citenamefont {Wagner},\ and\ \citenamefont {Norman}}]{K2013}%
  \BibitemOpen
  \bibfield  {author} {\bibinfo {author} {\bibfnamefont {A.~G.}\ \bibnamefont
  {Kritsuk}}, \bibinfo {author} {\bibfnamefont {R.}~\bibnamefont {Wagner}}, \
  and\ \bibinfo {author} {\bibfnamefont {M.~L.}\ \bibnamefont {Norman}},\
  }\href@noop {} {\bibfield  {journal} {\bibinfo  {journal} {Journal of Fluid
  Mechanics}\ }\textbf {\bibinfo {volume} {729}},\ \bibinfo {pages} {R1}
  (\bibinfo {year} {2013})}\BibitemShut {NoStop}%
\bibitem [{\citenamefont {G\'omez}\ \emph {et~al.}(2005)\citenamefont
  {G\'omez}, \citenamefont {Mininni},\ and\ \citenamefont {Dmitruk}}]{Go2005}%
  \BibitemOpen
  \bibfield  {author} {\bibinfo {author} {\bibfnamefont {D.~O.}\ \bibnamefont
  {G\'omez}}, \bibinfo {author} {\bibfnamefont {P.~D.}\ \bibnamefont
  {Mininni}}, \ and\ \bibinfo {author} {\bibfnamefont {P.}~\bibnamefont
  {Dmitruk}},\ }\href@noop {} {\bibfield  {journal} {\bibinfo  {journal} {Phys.
  Scripta T116}\ }\textbf {\bibinfo {volume} {123}} (\bibinfo {year}
  {2005})}\BibitemShut {NoStop}%
\bibitem [{\citenamefont {Mininni}\ \emph {et~al.}(2011)\citenamefont
  {Mininni}, \citenamefont {Rosenberg}, \citenamefont {Reddy},\ and\
  \citenamefont {Pouquet}}]{Mi2011}%
  \BibitemOpen
  \bibfield  {author} {\bibinfo {author} {\bibfnamefont {P.~D.}\ \bibnamefont
  {Mininni}}, \bibinfo {author} {\bibfnamefont {D.}~\bibnamefont {Rosenberg}},
  \bibinfo {author} {\bibfnamefont {R.}~\bibnamefont {Reddy}}, \ and\ \bibinfo
  {author} {\bibfnamefont {A.}~\bibnamefont {Pouquet}},\ }\href@noop {}
  {\bibfield  {journal} {\bibinfo  {journal} {Parallel Computing}\ }\textbf
  {\bibinfo {volume} {37}},\ \bibinfo {pages} {16} (\bibinfo {year}
  {2011})}\BibitemShut {NoStop}%
\bibitem [{\citenamefont {Ghosh}\ \emph {et~al.}(1993)\citenamefont {Ghosh},
  \citenamefont {Hossain},\ and\ \citenamefont {Matthaeus}}]{Gh1993}%
  \BibitemOpen
  \bibfield  {author} {\bibinfo {author} {\bibfnamefont {S.}~\bibnamefont
  {Ghosh}}, \bibinfo {author} {\bibfnamefont {M.}~\bibnamefont {Hossain}}, \
  and\ \bibinfo {author} {\bibfnamefont {W.~H.}\ \bibnamefont {Matthaeus}},\
  }\href {\doibase https://doi.org/10.1016/0010-4655(93)90103-J} {\bibfield
  {journal} {\bibinfo  {journal} {Computer Physics Communications}\ }\textbf
  {\bibinfo {volume} {74}},\ \bibinfo {pages} {18 } (\bibinfo {year}
  {1993})}\BibitemShut {NoStop}%
\bibitem [{\citenamefont {Dmitruk}\ \emph {et~al.}(2005)\citenamefont
  {Dmitruk}, \citenamefont {Matthaeus},\ and\ \citenamefont {Oughton}}]{D2005}%
  \BibitemOpen
  \bibfield  {author} {\bibinfo {author} {\bibfnamefont {P.}~\bibnamefont
  {Dmitruk}}, \bibinfo {author} {\bibfnamefont {W.~H.}\ \bibnamefont
  {Matthaeus}}, \ and\ \bibinfo {author} {\bibfnamefont {S.}~\bibnamefont
  {Oughton}},\ }\href {\doibase 10.1063/1.2128573} {\bibfield  {journal}
  {\bibinfo  {journal} {Physics of Plasmas}\ }\textbf {\bibinfo {volume}
  {12}},\ \bibinfo {pages} {112304} (\bibinfo {year} {2005})}\BibitemShut
  {NoStop}%
\bibitem [{\citenamefont {Taylor}\ \emph {et~al.}(2003)\citenamefont {Taylor},
  \citenamefont {Kurien},\ and\ \citenamefont {Eyink}}]{Ta2003}%
  \BibitemOpen
  \bibfield  {author} {\bibinfo {author} {\bibfnamefont {M.~A.}\ \bibnamefont
  {Taylor}}, \bibinfo {author} {\bibfnamefont {S.}~\bibnamefont {Kurien}}, \
  and\ \bibinfo {author} {\bibfnamefont {G.~L.}\ \bibnamefont {Eyink}},\
  }\href@noop {} {\bibfield  {journal} {\bibinfo  {journal} {Physical Review
  E}\ }\textbf {\bibinfo {volume} {68}},\ \bibinfo {pages} {026310} (\bibinfo
  {year} {2003})}\BibitemShut {NoStop}%
\bibitem [{\citenamefont {Martin}\ and\ \citenamefont
  {Mininni}(2010)}]{Ma2010}%
  \BibitemOpen
  \bibfield  {author} {\bibinfo {author} {\bibfnamefont {L.}~\bibnamefont
  {Martin}}\ and\ \bibinfo {author} {\bibfnamefont {P.}~\bibnamefont
  {Mininni}},\ }\href@noop {} {\bibfield  {journal} {\bibinfo  {journal}
  {Physical Review E}\ }\textbf {\bibinfo {volume} {81}},\ \bibinfo {pages}
  {016310} (\bibinfo {year} {2010})}\BibitemShut {NoStop}%
\bibitem [{\citenamefont {Arad}\ \emph {et~al.}(1999)\citenamefont {Arad},
  \citenamefont {Biferale}, \citenamefont {Mazzitelli},\ and\ \citenamefont
  {Procaccia}}]{A1999}%
  \BibitemOpen
  \bibfield  {author} {\bibinfo {author} {\bibfnamefont {I.}~\bibnamefont
  {Arad}}, \bibinfo {author} {\bibfnamefont {L.}~\bibnamefont {Biferale}},
  \bibinfo {author} {\bibfnamefont {I.}~\bibnamefont {Mazzitelli}}, \ and\
  \bibinfo {author} {\bibfnamefont {I.}~\bibnamefont {Procaccia}},\ }\href@noop
  {} {\bibfield  {journal} {\bibinfo  {journal} {Physical review letters}\
  }\textbf {\bibinfo {volume} {82}},\ \bibinfo {pages} {5040} (\bibinfo {year}
  {1999})}\BibitemShut {NoStop}%
\bibitem [{\citenamefont {Imazio}\ and\ \citenamefont
  {Mininni}(2017)}]{Im2017}%
  \BibitemOpen
  \bibfield  {author} {\bibinfo {author} {\bibfnamefont {P.~R.}\ \bibnamefont
  {Imazio}}\ and\ \bibinfo {author} {\bibfnamefont {P.}~\bibnamefont
  {Mininni}},\ }\href@noop {} {\bibfield  {journal} {\bibinfo  {journal}
  {Physical Review E}\ }\textbf {\bibinfo {volume} {95}},\ \bibinfo {pages}
  {033103} (\bibinfo {year} {2017})}\BibitemShut {NoStop}%
\bibitem [{\citenamefont {Kurien}\ and\ \citenamefont
  {Sreenivasan}(000a)}]{K2000a}%
  \BibitemOpen
  \bibfield  {author} {\bibinfo {author} {\bibfnamefont {S.}~\bibnamefont
  {Kurien}}\ and\ \bibinfo {author} {\bibfnamefont {K.~R.}\ \bibnamefont
  {Sreenivasan}},\ }\href@noop {} {\bibfield  {journal} {\bibinfo  {journal}
  {Physical Review E}\ }\textbf {\bibinfo {volume} {62}},\ \bibinfo {pages}
  {2206} (\bibinfo {year} {2000a})}\BibitemShut {NoStop}%
\bibitem [{\citenamefont {Kurien}\ \emph {et~al.}(000b)\citenamefont {Kurien},
  \citenamefont {L’vov}, \citenamefont {Procaccia},\ and\ \citenamefont
  {Sreenivasan}}]{K2000b}%
  \BibitemOpen
  \bibfield  {author} {\bibinfo {author} {\bibfnamefont {S.}~\bibnamefont
  {Kurien}}, \bibinfo {author} {\bibfnamefont {V.~S.}\ \bibnamefont {L’vov}},
  \bibinfo {author} {\bibfnamefont {I.}~\bibnamefont {Procaccia}}, \ and\
  \bibinfo {author} {\bibfnamefont {K.}~\bibnamefont {Sreenivasan}},\
  }\href@noop {} {\bibfield  {journal} {\bibinfo  {journal} {Physical Review
  E}\ }\textbf {\bibinfo {volume} {61}},\ \bibinfo {pages} {407} (\bibinfo
  {year} {2000b})}\BibitemShut {NoStop}%
\bibitem [{\citenamefont {Biferale}\ and\ \citenamefont
  {Toschi}(2001)}]{B2001}%
  \BibitemOpen
  \bibfield  {author} {\bibinfo {author} {\bibfnamefont {L.}~\bibnamefont
  {Biferale}}\ and\ \bibinfo {author} {\bibfnamefont {F.}~\bibnamefont
  {Toschi}},\ }\href@noop {} {\bibfield  {journal} {\bibinfo  {journal}
  {Physical review letters}\ }\textbf {\bibinfo {volume} {86}},\ \bibinfo
  {pages} {4831} (\bibinfo {year} {2001})}\BibitemShut {NoStop}%
\bibitem [{\citenamefont {Mininni}\ and\ \citenamefont
  {Pouquet}(2010)}]{Mi2010}%
  \BibitemOpen
  \bibfield  {author} {\bibinfo {author} {\bibfnamefont {P.~D.}\ \bibnamefont
  {Mininni}}\ and\ \bibinfo {author} {\bibfnamefont {A.}~\bibnamefont
  {Pouquet}},\ }\href@noop {} {\bibfield  {journal} {\bibinfo  {journal}
  {Physics of Fluids}\ }\textbf {\bibinfo {volume} {22}},\ \bibinfo {pages}
  {035105} (\bibinfo {year} {2010})}\BibitemShut {NoStop}%
\bibitem [{\citenamefont {Imazio}\ and\ \citenamefont
  {Mininni}(2011)}]{Im2011}%
  \BibitemOpen
  \bibfield  {author} {\bibinfo {author} {\bibfnamefont {P.~R.}\ \bibnamefont
  {Imazio}}\ and\ \bibinfo {author} {\bibfnamefont {P.}~\bibnamefont
  {Mininni}},\ }\href@noop {} {\bibfield  {journal} {\bibinfo  {journal}
  {Physical Review E}\ }\textbf {\bibinfo {volume} {83}},\ \bibinfo {pages}
  {066309} (\bibinfo {year} {2011})}\BibitemShut {NoStop}%
\bibitem [{\citenamefont {Matthaeus}\ and\ \citenamefont
  {Goldstein}(1982)}]{M1982}%
  \BibitemOpen
  \bibfield  {author} {\bibinfo {author} {\bibfnamefont {W.~H.}\ \bibnamefont
  {Matthaeus}}\ and\ \bibinfo {author} {\bibfnamefont {M.~L.}\ \bibnamefont
  {Goldstein}},\ }\href@noop {} {\bibfield  {journal} {\bibinfo  {journal} {J.
  Geophys. Res.}\ }\textbf {\bibinfo {volume} {87}},\ \bibinfo {pages} {6011}
  (\bibinfo {year} {1982})}\BibitemShut {NoStop}%
\bibitem [{\citenamefont {Federrath}(2016)}]{F2016}%
  \BibitemOpen
  \bibfield  {author} {\bibinfo {author} {\bibfnamefont {C.}~\bibnamefont
  {Federrath}},\ }\href {\doibase 10.1017/S0022377816001069} {\bibfield
  {journal} {\bibinfo  {journal} {Journal of Plasma Physics}\ }\textbf
  {\bibinfo {volume} {82}},\ \bibinfo {pages} {535820601} (\bibinfo {year}
  {2016})}\BibitemShut {NoStop}%
\bibitem [{\citenamefont {Passot}\ and\ \citenamefont
  {V\'azquez-Semadeni}(1998)}]{PV1998}%
  \BibitemOpen
  \bibfield  {author} {\bibinfo {author} {\bibfnamefont {T.}~\bibnamefont
  {Passot}}\ and\ \bibinfo {author} {\bibfnamefont {E.}~\bibnamefont
  {V\'azquez-Semadeni}},\ }\href {\doibase 10.1103/PhysRevE.58.4501} {\bibfield
   {journal} {\bibinfo  {journal} {Phys. Rev. E}\ }\textbf {\bibinfo {volume}
  {58}},\ \bibinfo {pages} {4501} (\bibinfo {year} {1998})}\BibitemShut
  {NoStop}%
\bibitem [{\citenamefont {{Federrath}}\ \emph {et~al.}(2010)\citenamefont
  {{Federrath}}, \citenamefont {{Roman-Duval}}, \citenamefont {{Klessen}},
  \citenamefont {{Schmidt}},\ and\ \citenamefont {{Mac Low}}}]{F2010}%
  \BibitemOpen
  \bibfield  {author} {\bibinfo {author} {\bibfnamefont {C.}~\bibnamefont
  {{Federrath}}}, \bibinfo {author} {\bibfnamefont {J.}~\bibnamefont
  {{Roman-Duval}}}, \bibinfo {author} {\bibfnamefont {R.~S.}\ \bibnamefont
  {{Klessen}}}, \bibinfo {author} {\bibfnamefont {W.}~\bibnamefont
  {{Schmidt}}}, \ and\ \bibinfo {author} {\bibfnamefont {M.-M.}\ \bibnamefont
  {{Mac Low}}},\ }\href {\doibase 10.1051/0004-6361/200912437} {\bibfield
  {journal} {\bibinfo  {journal} {A\&A}\ }\textbf {\bibinfo {volume} {512}},\
  \bibinfo {pages} {A81} (\bibinfo {year} {2010})}\BibitemShut {NoStop}%
\bibitem [{\citenamefont {Federrath}\ and\ \citenamefont
  {Banerjee}(2015)}]{FB2015}%
  \BibitemOpen
  \bibfield  {author} {\bibinfo {author} {\bibfnamefont {C.}~\bibnamefont
  {Federrath}}\ and\ \bibinfo {author} {\bibfnamefont {S.}~\bibnamefont
  {Banerjee}},\ }\href {\doibase 10.1093/mnras/stv180} {\bibfield  {journal}
  {\bibinfo  {journal} {Monthly Notices of the Royal Astronomical Society}\
  }\textbf {\bibinfo {volume} {448}},\ \bibinfo {pages} {3297} (\bibinfo {year}
  {2015})}\BibitemShut {NoStop}%
\bibitem [{\citenamefont {Nolan}\ \emph {et~al.}(2015)\citenamefont {Nolan},
  \citenamefont {Federrath},\ and\ \citenamefont {Sutherland}}]{N2015}%
  \BibitemOpen
  \bibfield  {author} {\bibinfo {author} {\bibfnamefont {C.~A.}\ \bibnamefont
  {Nolan}}, \bibinfo {author} {\bibfnamefont {C.}~\bibnamefont {Federrath}}, \
  and\ \bibinfo {author} {\bibfnamefont {R.~S.}\ \bibnamefont {Sutherland}},\
  }\href {\doibase 10.1093/mnras/stv1030} {\bibfield  {journal} {\bibinfo
  {journal} {Monthly Notices of the Royal Astronomical Society}\ }\textbf
  {\bibinfo {volume} {451}},\ \bibinfo {pages} {1380} (\bibinfo {year}
  {2015})}\BibitemShut {NoStop}%
\bibitem [{\citenamefont {Federrath}\ and\ \citenamefont
  {Klessen}(2012)}]{FK2012}%
  \BibitemOpen
  \bibfield  {author} {\bibinfo {author} {\bibfnamefont {C.}~\bibnamefont
  {Federrath}}\ and\ \bibinfo {author} {\bibfnamefont {R.~S.}\ \bibnamefont
  {Klessen}},\ }\href {http://stacks.iop.org/0004-637X/761/i=2/a=156}
  {\bibfield  {journal} {\bibinfo  {journal} {The Astrophysical Journal}\
  }\textbf {\bibinfo {volume} {761}},\ \bibinfo {pages} {156} (\bibinfo {year}
  {2012})}\BibitemShut {NoStop}%
\bibitem [{\citenamefont {{Federrath}}\ \emph {et~al.}(2017)\citenamefont
  {{Federrath}}, \citenamefont {{Rathborne}}, \citenamefont {{Longmore}},
  \citenamefont {{Kruijssen}}, \citenamefont {{Bally}}, \citenamefont
  {{Contreras}}, \citenamefont {{Crocker}}, \citenamefont {{Garay}},
  \citenamefont {{Jackson}}, \citenamefont {{Testi}},\ and\ \citenamefont
  {{Walsh}}}]{F2017}%
  \BibitemOpen
  \bibfield  {author} {\bibinfo {author} {\bibfnamefont {C.}~\bibnamefont
  {{Federrath}}}, \bibinfo {author} {\bibfnamefont {J.~M.}\ \bibnamefont
  {{Rathborne}}}, \bibinfo {author} {\bibfnamefont {S.~N.}\ \bibnamefont
  {{Longmore}}}, \bibinfo {author} {\bibfnamefont {J.~M.~D.}\ \bibnamefont
  {{Kruijssen}}}, \bibinfo {author} {\bibfnamefont {J.}~\bibnamefont
  {{Bally}}}, \bibinfo {author} {\bibfnamefont {Y.}~\bibnamefont
  {{Contreras}}}, \bibinfo {author} {\bibfnamefont {R.~M.}\ \bibnamefont
  {{Crocker}}}, \bibinfo {author} {\bibfnamefont {G.}~\bibnamefont {{Garay}}},
  \bibinfo {author} {\bibfnamefont {J.~M.}\ \bibnamefont {{Jackson}}}, \bibinfo
  {author} {\bibfnamefont {L.}~\bibnamefont {{Testi}}}, \ and\ \bibinfo
  {author} {\bibfnamefont {A.~J.}\ \bibnamefont {{Walsh}}},\ }in\ \href
  {\doibase 10.1017/S1743921316012357} {\emph {\bibinfo {booktitle} {The
  Multi-Messenger Astrophysics of the Galactic Centre}}},\ \bibinfo {series}
  {IAU Symposium}, Vol.\ \bibinfo {volume} {322},\ \bibinfo {editor} {edited
  by\ \bibinfo {editor} {\bibfnamefont {R.~M.}\ \bibnamefont {{Crocker}}},
  \bibinfo {editor} {\bibfnamefont {S.~N.}\ \bibnamefont {{Longmore}}}, \ and\
  \bibinfo {editor} {\bibfnamefont {G.~V.}\ \bibnamefont {{Bicknell}}}}\
  (\bibinfo {year} {2017})\ pp.\ \bibinfo {pages} {123--128}\BibitemShut
  {NoStop}%
\bibitem [{\citenamefont {Shebalin}\ \emph {et~al.}(1983)\citenamefont
  {Shebalin}, \citenamefont {Matthaeus},\ and\ \citenamefont
  {Montgomery}}]{Sh1983}%
  \BibitemOpen
  \bibfield  {author} {\bibinfo {author} {\bibfnamefont {J.~V.}\ \bibnamefont
  {Shebalin}}, \bibinfo {author} {\bibfnamefont {W.~H.}\ \bibnamefont
  {Matthaeus}}, \ and\ \bibinfo {author} {\bibfnamefont {D.}~\bibnamefont
  {Montgomery}},\ }\href@noop {} {\bibfield  {journal} {\bibinfo  {journal}
  {Journal of Plasma Physics}\ }\textbf {\bibinfo {volume} {29}},\ \bibinfo
  {pages} {525} (\bibinfo {year} {1983})}\BibitemShut {NoStop}%
\bibitem [{\citenamefont {Matthaeus}\ \emph {et~al.}(1996)\citenamefont
  {Matthaeus}, \citenamefont {Ghosh}, \citenamefont {Oughton},\ and\
  \citenamefont {Roberts}}]{M1996}%
  \BibitemOpen
  \bibfield  {author} {\bibinfo {author} {\bibfnamefont {W.~H.}\ \bibnamefont
  {Matthaeus}}, \bibinfo {author} {\bibfnamefont {S.}~\bibnamefont {Ghosh}},
  \bibinfo {author} {\bibfnamefont {S.}~\bibnamefont {Oughton}}, \ and\
  \bibinfo {author} {\bibfnamefont {D.~A.}\ \bibnamefont {Roberts}},\
  }\href@noop {} {\bibfield  {journal} {\bibinfo  {journal} {Journal of
  Geophysical Research: Space Physics}\ }\textbf {\bibinfo {volume} {101}},\
  \bibinfo {pages} {7619} (\bibinfo {year} {1996})}\BibitemShut {NoStop}%
\bibitem [{\citenamefont {Low}(1999)}]{Ma1999}%
  \BibitemOpen
  \bibfield  {author} {\bibinfo {author} {\bibfnamefont {M.-M.~M.}\
  \bibnamefont {Low}},\ }\href {http://stacks.iop.org/0004-637X/524/i=1/a=169}
  {\bibfield  {journal} {\bibinfo  {journal} {The Astrophysical Journal}\
  }\textbf {\bibinfo {volume} {524}},\ \bibinfo {pages} {169} (\bibinfo {year}
  {1999})}\BibitemShut {NoStop}%
\bibitem [{\citenamefont {Galtier}\ \emph {et~al.}(2000)\citenamefont
  {Galtier}, \citenamefont {Nazarenko}, \citenamefont {Newell},\ and\
  \citenamefont {Pouquet}}]{Ga2000}%
  \BibitemOpen
  \bibfield  {author} {\bibinfo {author} {\bibfnamefont {S.}~\bibnamefont
  {Galtier}}, \bibinfo {author} {\bibfnamefont {S.~V.}\ \bibnamefont
  {Nazarenko}}, \bibinfo {author} {\bibfnamefont {A.~C.}\ \bibnamefont
  {Newell}}, \ and\ \bibinfo {author} {\bibfnamefont {A.}~\bibnamefont
  {Pouquet}},\ }\href@noop {} {\bibfield  {journal} {\bibinfo  {journal}
  {Journal of Plasma Physics}\ }\textbf {\bibinfo {volume} {63}},\ \bibinfo
  {pages} {447–488} (\bibinfo {year} {2000})}\BibitemShut {NoStop}%
\bibitem [{\citenamefont {Brunt}\ \emph {et~al.}(2010)\citenamefont {Brunt},
  \citenamefont {Federrath},\ and\ \citenamefont {Price}}]{Br2010}%
  \BibitemOpen
  \bibfield  {author} {\bibinfo {author} {\bibfnamefont {C.~M.}\ \bibnamefont
  {Brunt}}, \bibinfo {author} {\bibfnamefont {C.}~\bibnamefont {Federrath}}, \
  and\ \bibinfo {author} {\bibfnamefont {D.~J.}\ \bibnamefont {Price}},\ }\href
  {\doibase 10.1111/j.1365-2966.2009.16215.x} {\bibfield  {journal} {\bibinfo
  {journal} {Monthly Notices of the Royal Astronomical Society}\ }\textbf
  {\bibinfo {volume} {403}},\ \bibinfo {pages} {1507} (\bibinfo {year}
  {2010})}\BibitemShut {NoStop}%
\bibitem [{\citenamefont {Wan}\ \emph {et~al.}(2012)\citenamefont {Wan},
  \citenamefont {Oughton}, \citenamefont {Servidio},\ and\ \citenamefont
  {Matthaeus}}]{W2012}%
  \BibitemOpen
  \bibfield  {author} {\bibinfo {author} {\bibfnamefont {M.}~\bibnamefont
  {Wan}}, \bibinfo {author} {\bibfnamefont {S.}~\bibnamefont {Oughton}},
  \bibinfo {author} {\bibfnamefont {S.}~\bibnamefont {Servidio}}, \ and\
  \bibinfo {author} {\bibfnamefont {W.~H.}\ \bibnamefont {Matthaeus}},\
  }\href@noop {} {\bibfield  {journal} {\bibinfo  {journal} {J. Fluid Mech.}\
  }\textbf {\bibinfo {volume} {697}},\ \bibinfo {pages} {296} (\bibinfo {year}
  {2012})}\BibitemShut {NoStop}%
\bibitem [{\citenamefont {Oughton}\ \emph {et~al.}(2013)\citenamefont
  {Oughton}, \citenamefont {Wan}, \citenamefont {Servidio},\ and\ \citenamefont
  {Matthaeus}}]{Ou2013}%
  \BibitemOpen
  \bibfield  {author} {\bibinfo {author} {\bibfnamefont {S.}~\bibnamefont
  {Oughton}}, \bibinfo {author} {\bibfnamefont {M.}~\bibnamefont {Wan}},
  \bibinfo {author} {\bibfnamefont {S.}~\bibnamefont {Servidio}}, \ and\
  \bibinfo {author} {\bibfnamefont {W.~H.}\ \bibnamefont {Matthaeus}},\ }\href
  {http://stacks.iop.org/0004-637X/768/i=1/a=10} {\bibfield  {journal}
  {\bibinfo  {journal} {The Astrophysical Journal}\ }\textbf {\bibinfo {volume}
  {768}},\ \bibinfo {pages} {10} (\bibinfo {year} {2013})}\BibitemShut
  {NoStop}%
\bibitem [{\citenamefont {Meyrand}\ \emph {et~al.}(2016)\citenamefont
  {Meyrand}, \citenamefont {Galtier},\ and\ \citenamefont {Kiyani}}]{Me2016}%
  \BibitemOpen
  \bibfield  {author} {\bibinfo {author} {\bibfnamefont {R.}~\bibnamefont
  {Meyrand}}, \bibinfo {author} {\bibfnamefont {S.}~\bibnamefont {Galtier}}, \
  and\ \bibinfo {author} {\bibfnamefont {K.~H.}\ \bibnamefont {Kiyani}},\
  }\href@noop {} {\bibfield  {journal} {\bibinfo  {journal} {Phys. Rev. Lett.}\
  }\textbf {\bibinfo {volume} {116}},\ \bibinfo {pages} {105002} (\bibinfo
  {year} {2016})}\BibitemShut {NoStop}%
\bibitem [{\citenamefont {{Oughton}}\ \emph {et~al.}(2016)\citenamefont
  {{Oughton}}, \citenamefont {{Matthaeus}}, \citenamefont {{Wan}},\ and\
  \citenamefont {{Parashar}}}]{Ou2016}%
  \BibitemOpen
  \bibfield  {author} {\bibinfo {author} {\bibfnamefont {S.}~\bibnamefont
  {{Oughton}}}, \bibinfo {author} {\bibfnamefont {W.~H.}\ \bibnamefont
  {{Matthaeus}}}, \bibinfo {author} {\bibfnamefont {M.}~\bibnamefont {{Wan}}},
  \ and\ \bibinfo {author} {\bibfnamefont {T.}~\bibnamefont {{Parashar}}},\
  }\href@noop {} {\bibfield  {journal} {\bibinfo  {journal} {J. Geophys. Res.}\
  }\textbf {\bibinfo {volume} {121}},\ \bibinfo {pages} {5041} (\bibinfo {year}
  {2016})}\BibitemShut {NoStop}%
\bibitem [{\citenamefont {Sujovolsky}\ and\ \citenamefont
  {Mininni}(2016)}]{S2016}%
  \BibitemOpen
  \bibfield  {author} {\bibinfo {author} {\bibfnamefont {N.~E.}\ \bibnamefont
  {Sujovolsky}}\ and\ \bibinfo {author} {\bibfnamefont {P.~D.}\ \bibnamefont
  {Mininni}},\ }\href {\doibase 10.1103/PhysRevFluids.1.054407} {\bibfield
  {journal} {\bibinfo  {journal} {Phys. Rev. Fluids}\ }\textbf {\bibinfo
  {volume} {1}},\ \bibinfo {pages} {054407} (\bibinfo {year}
  {2016})}\BibitemShut {NoStop}%
\end{thebibliography}

%merlin.mbs apsrev4-1.bst 2010-07-25 4.21a (PWD, AO, DPC) hacked
%Control: key (0)
%Control: author (72) initials jnrlst
%Control: editor formatted (1) identically to author
%Control: production of article title (-1) disabled
%Control: page (0) single
%Control: year (1) truncated
%Control: production of eprint (0) enabled
%

\end{document}